\documentclass[a4paper,11pt]{article}
\usepackage{pos}
\renewcommand{\logo}{\relax}

\title{The disc instability model: original recipe and additional ingredients}
\ShortTitle{The DIM: original recipe and additional ingredients}

\author*{J.M. Hameury}

\affiliation{Universit\'e de Strasbourg, CNRS, Observatoire astronomique de Strasbourg, UMR7550, 67000 Strasbourg, France}

\emailAdd{jean-marie.hameury@astro.unistra.fr}

\abstract{The disc instability model successfully reproduces many of the observed properties of cataclysmic variables. However, additional ingredients such as mass-transfer variations, disc irradiation, stream-disc overflow, or inner-disc truncation must be included to explain certain systems. The physics underlying these processes is often poorly constrained, and our lack of knowledge is typically absorbed into extra free parameters, much like the $\alpha$-prescription for viscosity. In this paper, I examine how each of these ingredients affects the predicted light curves and discuss the limitations that arise from the growing number of unconstrained parameters on the model's predictive power.}

\FullConference{87th Fujihara Seminar: The 50th Anniversary Workshop of the disc Instability Model in Compact Binary Stars (DIM50TH2025) \\
22-26 September 2025\\
Tomakomai, Japan \\}

\begin{document}
\maketitle

\section{Introduction}
The disc instability model (DIM) was proposed 50 years ago \cite{o74} to explain the outbursts observed in dwarf novae (DNe). These systems form a subclass of cataclysmic variables, i.e., binaries in which a white dwarf accretes matter from a companion through Roche lobe overflow. Dwarf novae display regular outbursts lasting typically a few days, recurring  on timescales of several weeks. Extended reviews of the model can be found in \cite{l01,h20}.

In the DIM, the accretion disc can locally be either in a cold state, with effective temperatures $T_{\rm eff}$ below $\sim$ 5,800 K, low viscosity and hence a low mass flow rate, or in a hot state, with $T_{\rm eff} > 7,200$ K, high viscosity and correspondingly large mass transfer rates. Between these two stable branches lies an intermediate, thermally and viscously unstable branch. If the mass transfer rate from the secondary causes any region of the disc to have $T_{\rm eff}$ within the range 5,800 -- 7,200 K, the disc cannot remain steady and instead undergoes a limit-cycle, producing the observed outbursts.

Because the DIM successfully reproduces the main characteristics of DN outbursts, such as their amplitude, duration, and recurrence time, it has become the standard explanation for these systems. With some modifications, the model has also been applied to soft X-ray transients (SXTs), which are closely related systems where the accreting object is a neutron star or a black hole. In these cases, the outbursts are much longer and the recurrence times significantly larger. 

The DIM however relies on several simplifying assumptions, many of which are at best questionable. In its original form, it also fails to reproduce the outbursts of soft X-ray transients and certain peculiar subclasses of dwarf novae. These shortcomings have motivated the introduction of additional physical ingredients. In what follows, I first outline the original formulation of the DIM and then present the extensions that have been developed to overcome its limitations, discussing in each case their effects on the predicted light curves and outburst properties. To facilitate comparison and highlight the influence of each modification, the same binary parameters are adopted throughout: a primary mass of $M_1 = 0.6$ ~M$_\odot$, an orbital period $P_{\mathrm{orb}}$ of 4~hr, and identical viscosity parameters of $\alpha_{\mathrm{c}} = 0.04$ and $\alpha_{\mathrm{h}} = 0.2$, unless stated otherwise, even when this choice may appear somewhat artificial.

The numerical code used here is described in \cite{hmd98}. It is a 1D code that solves the mass, angular momentum and energy conservation equations using an adaptive grid technique and an implicit numerical scheme, which allows transition fronts to be accurately resolved. The vertical structures are pre-computed on a large grid ($21\times 100 \times 230 \times 350$ in irradiation temperatures, radius, surface density, and mid-plane temperatures respectively) and interpolated to evaluate the cooling term in the energy equation. Both the inner and outer disc radii are allowed to vary with time; a tidal torque (see section 5.3 for details) is included and limits the radial extent of the disc. Time derivatives are computed using a Lagrangian scheme. The surface density is set to zero at the inner edge and, unless otherwise stated, mass is added at the outer disc edge. The grid typically consists of 1000 points. As for the initial conditions, one starts from a disc on the hot branch with a high mass-transfer rate, which is then reduced to the desired value. The code is run until a relaxed state is reached and the light curve is periodic.

The light curves shown here include contributions from the disc, the white dwarf, and the hot spot, but not from the secondary star. Unless otherwise stated, the white dwarf temperature is set to zero. The distance is assumed to be 109 pc, and the inclination angle is fixed at 30$^\circ$.

\section{The original model}

The DIM assumes that the accretion disc is geometrically thin, optically thick and axisymmetric. The first two assumptions are generally well justified, although the optically thick assumption may break down in the innermost regions during low states \cite{ilh10}. The geometrically thin assumption is crucial as it allows the vertical and radial structures of the disc to be decoupled, thereby reducing the problem to an effectively one-dimensional treatment.
The assumption of axisymmetry is, however, clearly an oversimplification. The presence of the donor star introduces strong tidal forces that truncate the disc at a finite radius. To handle this, one must either assume that the outer disc radius remains fixed at a prescribed value or include a tidal torque term, $\mathcal{T}(r)$, which constrains the disc’s radial extent and enable its variations over time. Other consequences of the axisymmetric assumption are that spiral density waves are neglected and that tidal instabilities, such as that proposed by Osaki \cite{o89} to explain the SU UMa phenomenon, have to be introduced {\it ad hoc}.

The most important assumption concerns the transport of angular momentum, which is assumed to be due to viscosity described by the $\alpha$ parametrization \cite{ss73}. Whereas the $\alpha$ parametrization is not a serious drawback, because $\alpha$ may depend on various parameters -- and this is what one does when using a different value of $\alpha$ in outburst and in quiescence, assuming that angular momentum transport is described by a viscous process means that angular momentum transport and energy dissipation are coupled and are local processes. This would not be the case if spiral waves play an important role.

Several additional assumptions are also commonly made for convenience, though they are not intrinsic to the DIM and can, in principle, be relaxed. These include, for example, the assumption that mass is added to the disc exclusively at its outer edge, and that the $\alpha$-prescription applies uniformly in the vertical direction, i.e., that the viscous energy dissipation rate is proportional to the total pressure throughout the disc.

Using these assumptions, one can solve the vertical structure equations of the disc and obtain the well-known S-shaped thermal equilibrium curve (often called the ``S-curve") which relates the local surface density of the disc and the effective disc temperature (or equivalently the local mass flow rate). The upper and lower branches are stable. The intermediate one is unstable and exists because of the steep dependence of opacities as a function of temperature. Outbursts occur when, due to mass accumulation in the disc, the surface density $\Sigma$ at some radius exceeds the critical value $\Sigma_{\mathrm{max}}$ corresponding to the upper limit of the cold branch. A heating front is then triggered and propagates through the disc, switching it to the hot state. As accretion proceeds, the disc mass decreases until $\Sigma$ drops below $\Sigma_{\mathrm{min}}$, the lower limit of the hot branch, at which point a cooling front starts and the disc returns to quiescence. 

\subsection{Viscosity}

\begin{figure}
\begin{center}
\includegraphics[width=0.33\columnwidth]{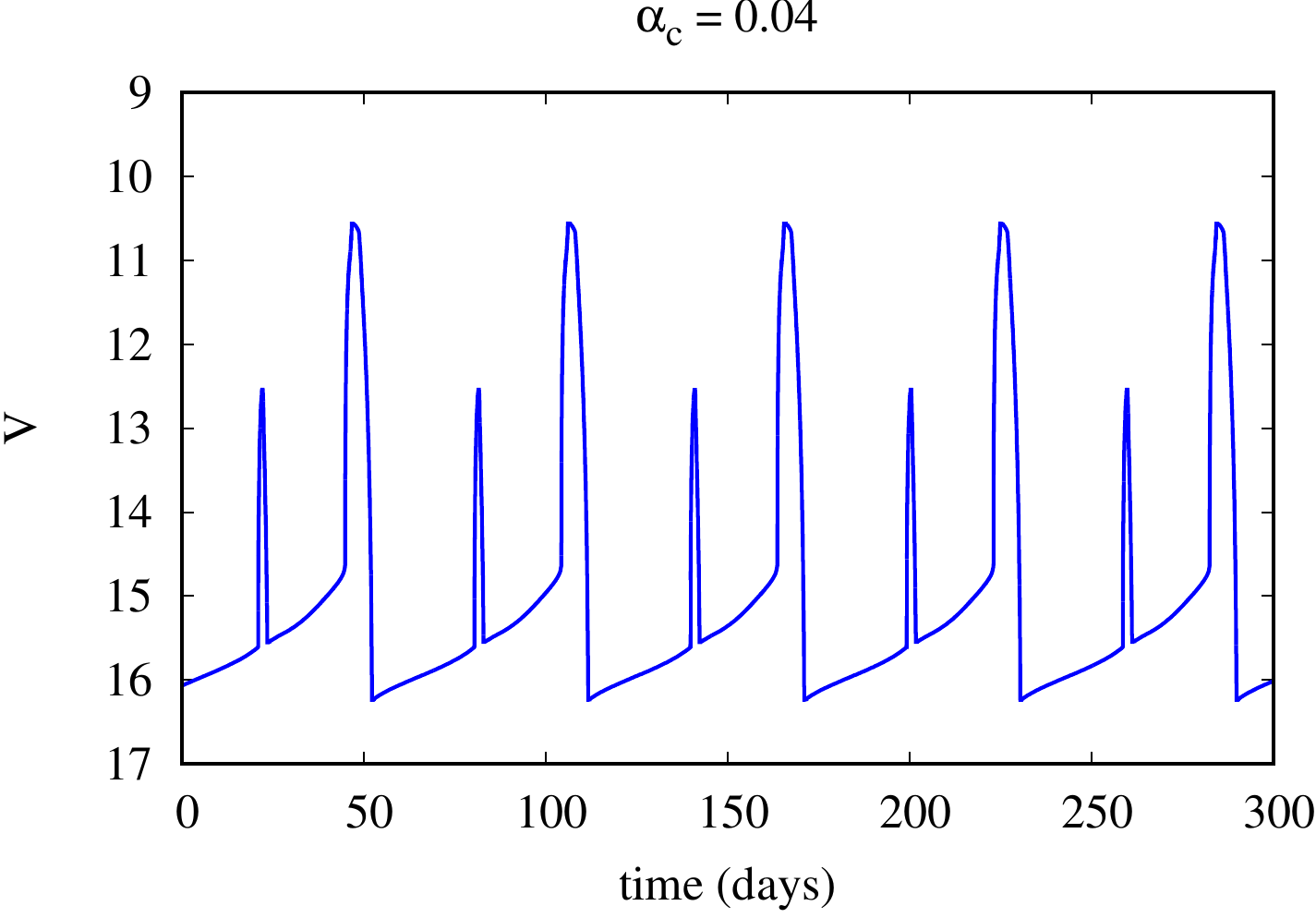}\includegraphics[width=0.33\columnwidth]{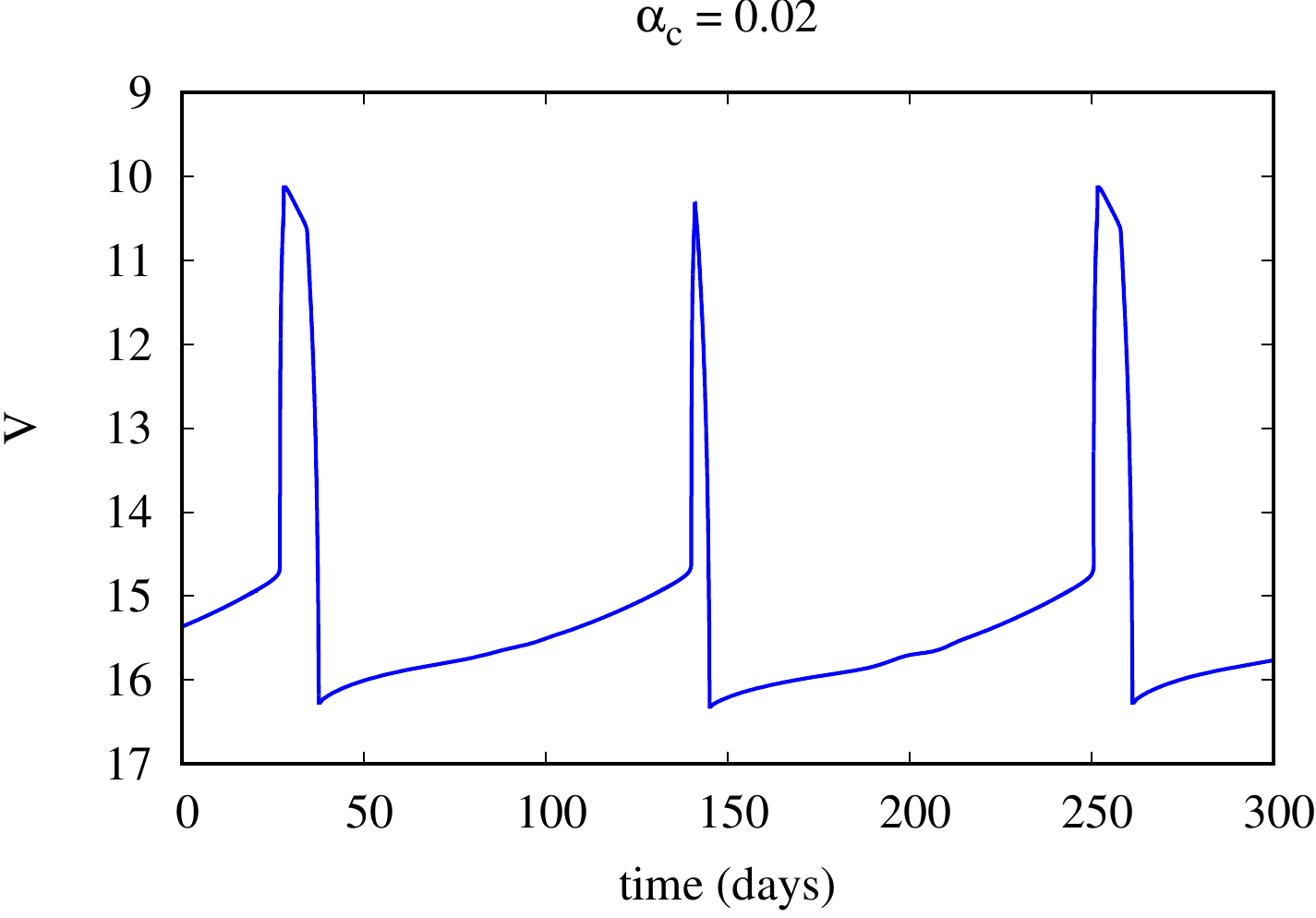}\includegraphics[width=0.33\columnwidth]{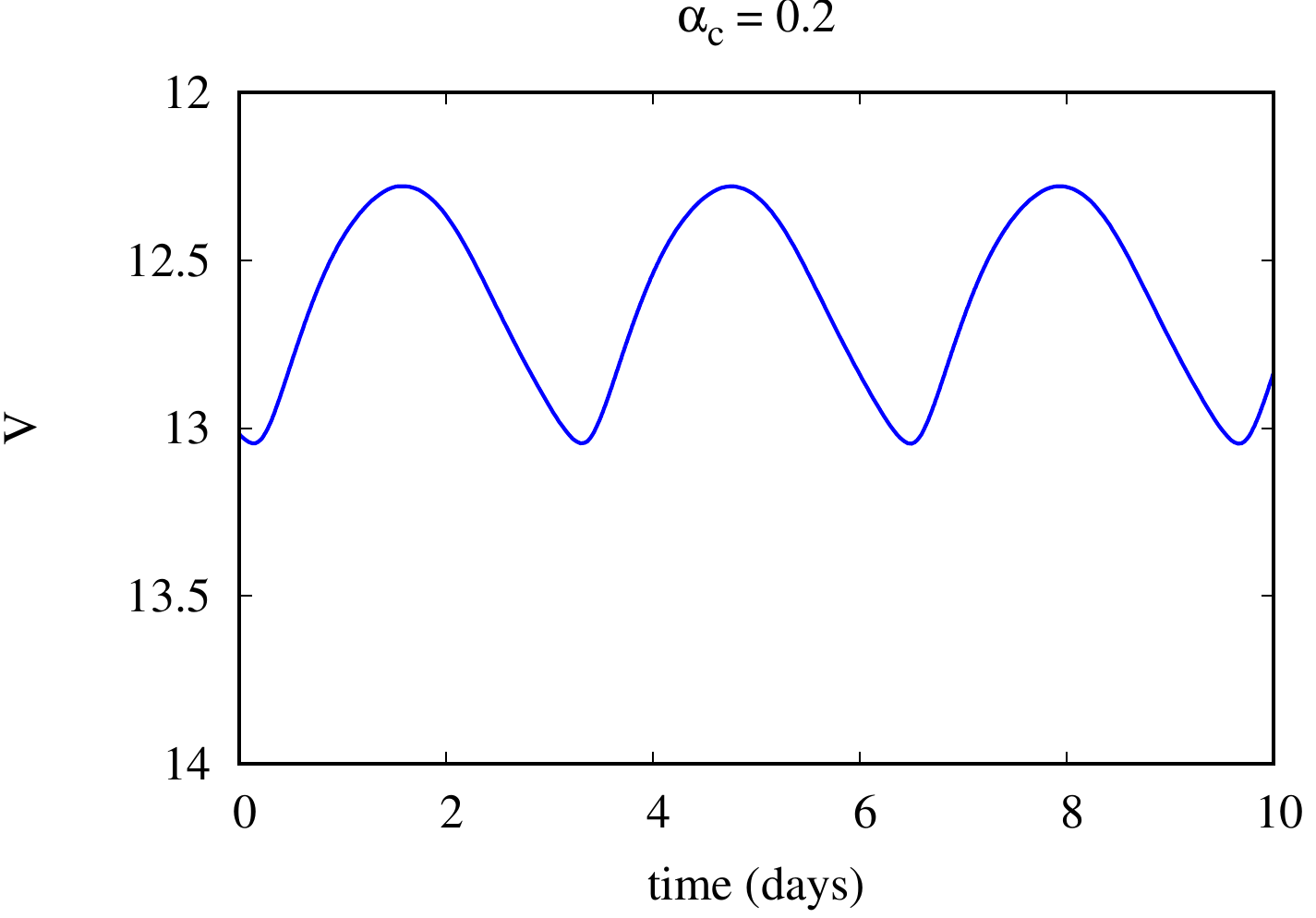}
\end{center}
\caption{Influence of $\alpha_{\rm c}$ on the light curve. The primary mass is $M_1=0.6$ M$_\odot$, the orbital period is 4 hr, and the viscosity parameter is $\alpha_h=0.2$ on the hot branch.}
\label{fig:visc}
\end{figure}

The impact of the viscosity parameter(s) on the outburst properties has been studied since the early days of the DIM. It was first realized that the simplest $\alpha = Const.$ hypothesis produces only small luminosity oscillations rather than full-scale outbursts \cite{s84} (see Fig. \ref{fig:visc}). Over the years, various prescriptions have been explored, including dependencies on the disc's aspect ratio, $(H/r)^n$, where $H$ is the vertical scale height , or on the radial coordinate alone, $r^\epsilon$ \cite{c94,ccl95}. Most modern implementations now adopt two different constant values of $\alpha$ on the cold and hot branches, with a smooth but steep transition between the two. Typical values of $\alpha = \alpha_{\rm h} \sim 0.1 - 0.2$ are used on the hot branch. These are relatively well constrained by the outburst durations \cite{kl12}, since larger $\alpha$ values yield faster accretion, and therefore shorter decay phases and overall shorter outbursts. The situation is far less certain for the cold branch, as  the physical mechanism responsible for sustaining viscosity at low temperatures -- where the magnetorotational instability (MRI) is expected to be inefficient -- remains unclear.

To reproduce observed outburst cycles, a viscosity contrast of $\alpha_{\rm h}/\alpha_{\rm c} \sim 5$–$10$ is generally required. Small values of $\alpha_{\rm c}$ produce long recurrence times; indeed, $\alpha_{\rm c} \sim 10^{-3}$ has been proposed to account for the extended quiescent intervals observed in soft X-ray transients (SXTs) \cite{mhl00} and WZ Sge-type systems \cite{o95}. A functional dependence such as $\alpha \propto (H/r)^n$ with $n = 2$ \cite{ccl95} also yields low $\alpha$ values in quiescence. Figure \ref{fig:visc} illustrates the effect of varying $\alpha_{\rm h}/\alpha_{\rm c}$ on the predicted light curves.

\section{Disc truncation}

\begin{figure}
\begin{center}
\includegraphics[width=0.45\columnwidth]{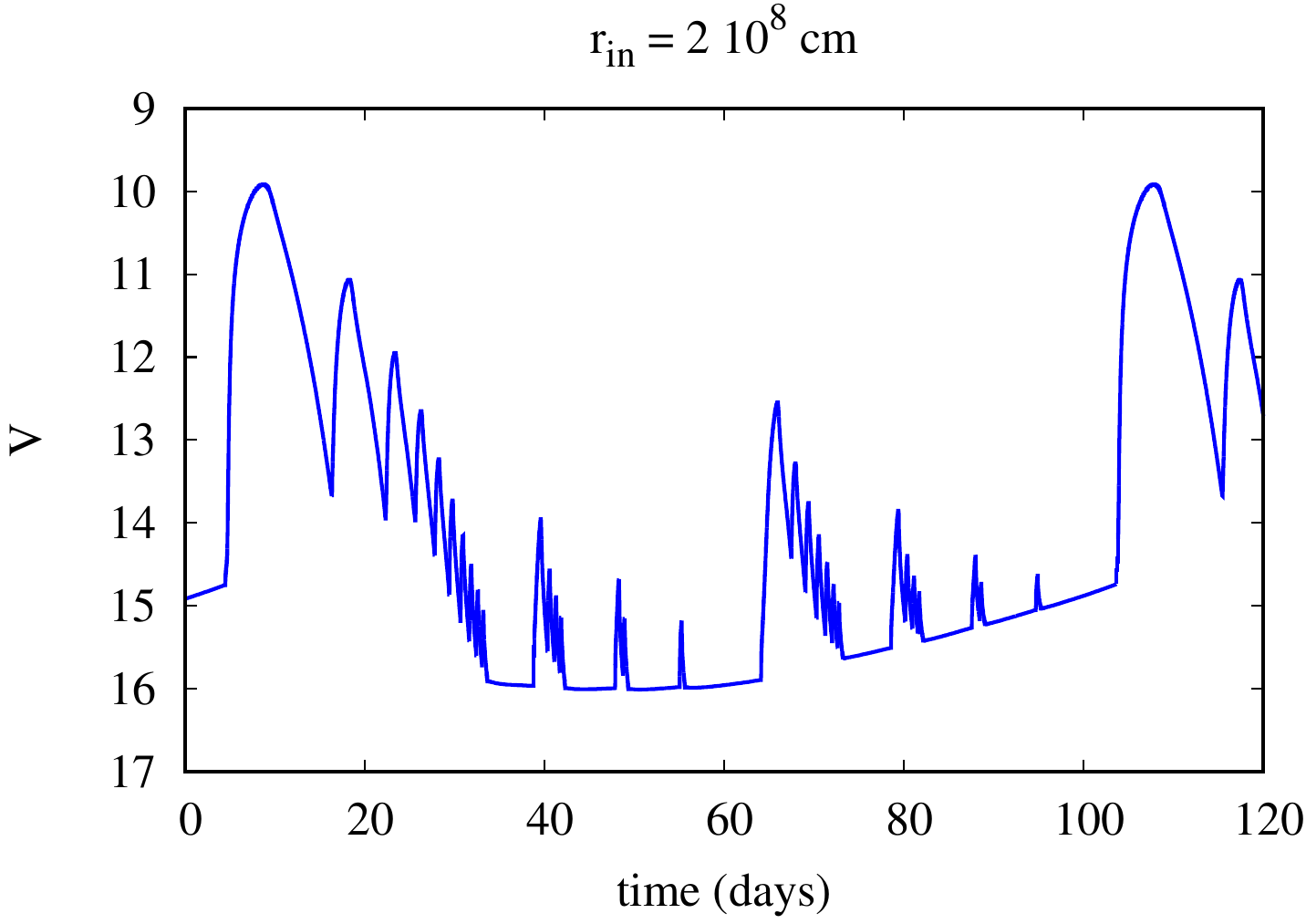}\includegraphics[width=0.45\columnwidth]{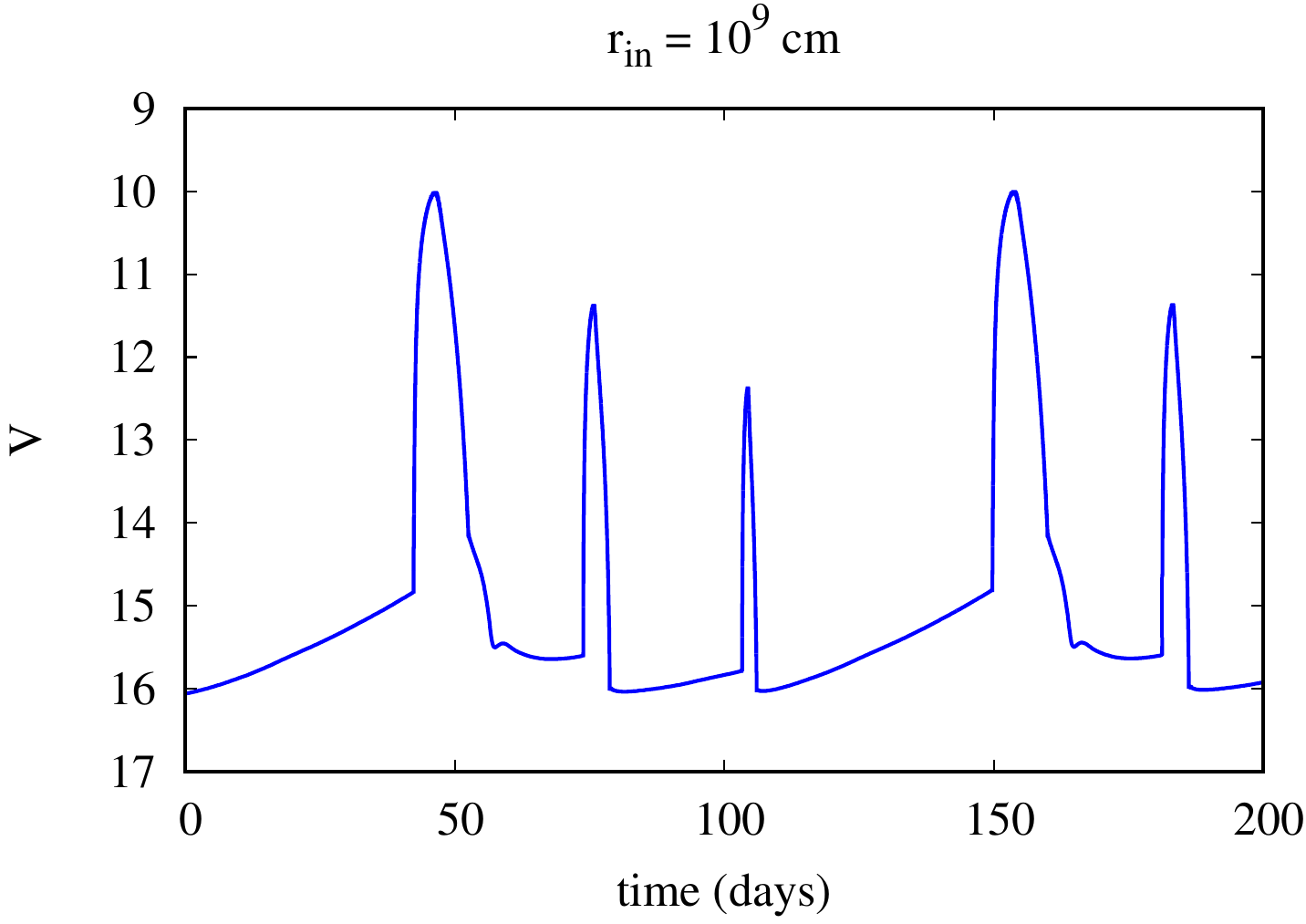}
\includegraphics[width=0.45\columnwidth]{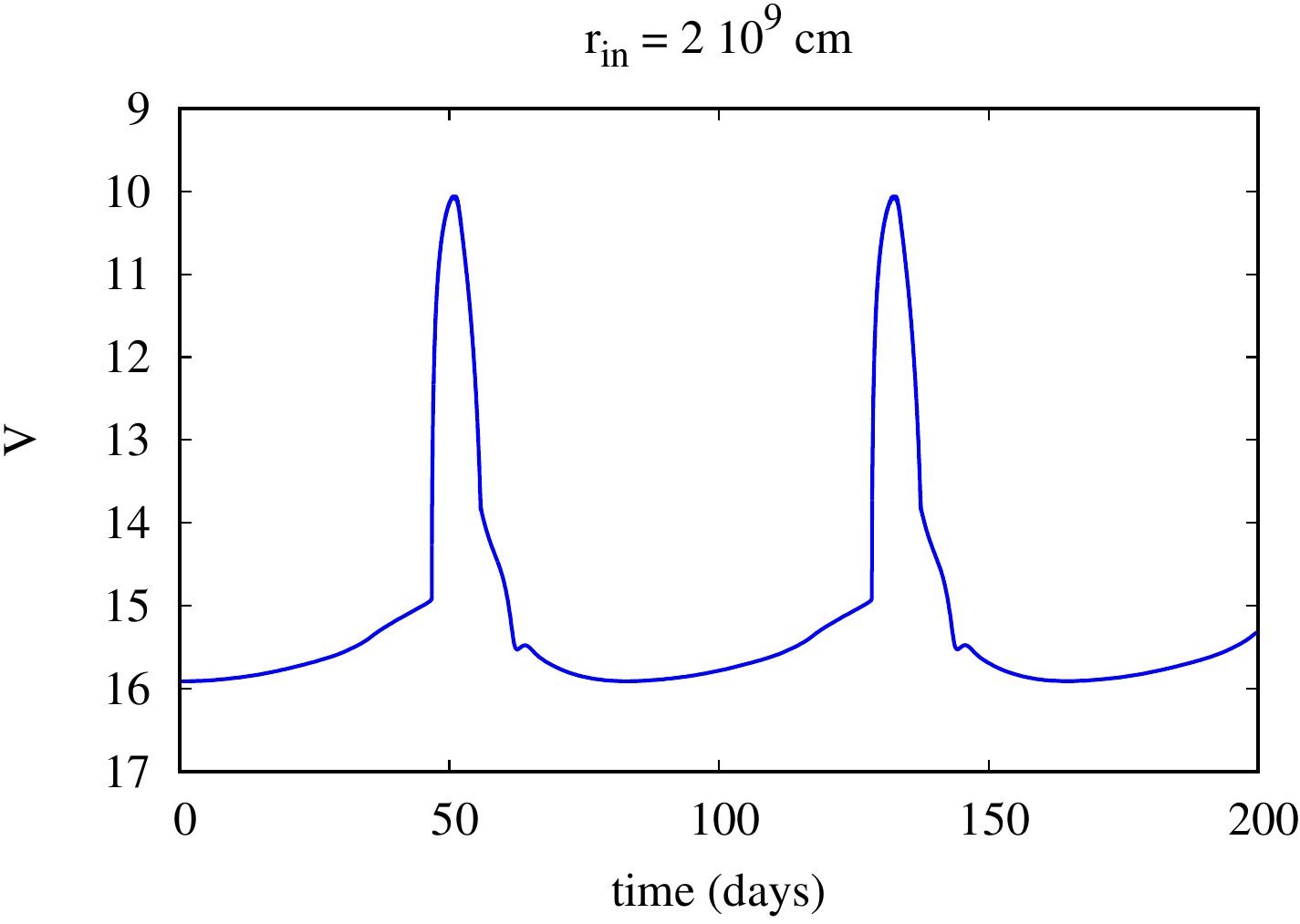}\includegraphics[width=0.45\columnwidth]{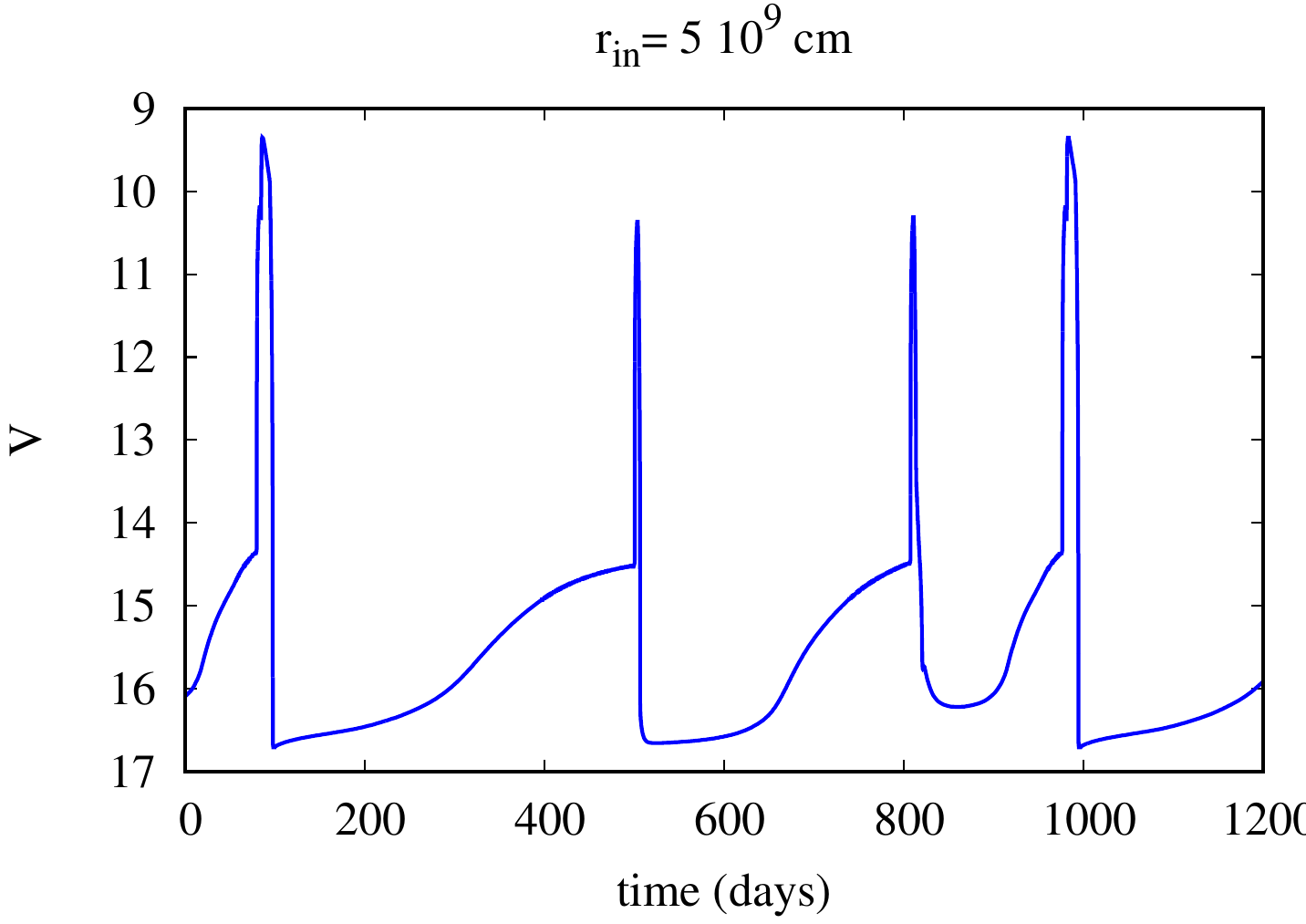}
\end{center}
\caption{Influence of the truncation of the inner disc on the light curve. The primary mass is $M_1=1$M$_\odot$, the orbital period is 4 hr; $\alpha_c=0.04$ on the cold branch and $\alpha_h=0.2$ on the hot branch.}
\label{fig:rin}
\end{figure}

When applied to systems with high primary masses and long orbital periods, such as black hole binaries, the model predicts the occurrence of multiple reflares during the decline phase of an outburst. These arise because, although the surface density $\Sigma$ is below the critical value $\Sigma_{\rm max}$ on the cold branch after the passage of a cooling front, the ratio $\Sigma/\Sigma_{\rm max}$ gradually increases toward smaller radii and may reach unity if the inner disc radius is sufficiently small. At this point, the cooling front is reflected into a heating front, thereby initiating a reflare. 

Although reflares are frequently observed in SXTs, the predicted ones differ markedly from the observed ones, as illustrated in Fig. \ref{fig:rin}. This figure presents light curves for a system with orbital parameters typical of a neutron star SXT. In this case, dozens of reflares appear during the decay phase when the inner radius is $2\times10^8$ cm, already much larger than the neutron star radius. Furthermore, numerous short, weak outbursts occur between two major outbursts, and the recurrence time between successive major events is unrealistically short, on the order of three months.

Since the origin of the reflares lies in the increase of $\Sigma/\Sigma_{\rm max}$ toward smaller radii, an obvious remedy is to enlarge the inner disc radius. This can occur either due to the presence of a magnetic field -- possible when the accretor is a white dwarf or a neutron star -- or because the accretion flow changes character near the compact object, forming, for example, an advection-dominated accretion flow (ADAF). As shown in Fig. \ref{fig:rin}, increasing $r_{\rm in}$ to $10^9$ cm is sufficient to suppress the reflares, although weak, intermediate outbursts persist. When $r_{\rm in}$ is increased to $2\times10^9$ cm, even these intermediate outbursts disappear. Interestingly, they reappear if $r_{\rm in}$ is further increased to $5\times10^9$ cm. In this configuration, the system is nearly stable: the accretion rate just before an outburst onset is roughly one-third of the mass-transfer rate. The major outbursts are of the outside-in type, i.e. triggered at large distances in the disc, lasting about 18 days, with recurrence times close to three years. Even in this extreme case, however, the recurrence time -- particularly when intermediate outbursts are considered -- remains shorter than observed values (years for neutron star SXTs and decades for black hole systems), indicating that additional physical effects must be taken into account.

 \section{Disc irradiation}

Irradiation of the accretion disc alters its internal structure by introducing an additional term in the boundary condition for the vertical structure equations:
\begin{equation}
\sigma T^4 \left( \tau = \frac{2}{3} \right) = F + \sigma T_{\rm irr}^4
\end{equation}
provided that the irradiating flux, $\sigma T_{\rm irr}^4$, can penetrate below the disc photosphere. Here, $F$ denotes the internal energy flux within the disc.

Irradiation substantially modifies the disc’s stability properties once $T_{\rm irr}$ becomes comparable to the maximum temperature on the cold branch. For irradiation temperatures exceeding approximately 10,000 K, the cold and intermediate branches of the thermal equilibrium curve vanish entirely, leaving only the hot, thermally stable configuration.

\begin{figure}
\begin{center}
\includegraphics[width=0.33\columnwidth]{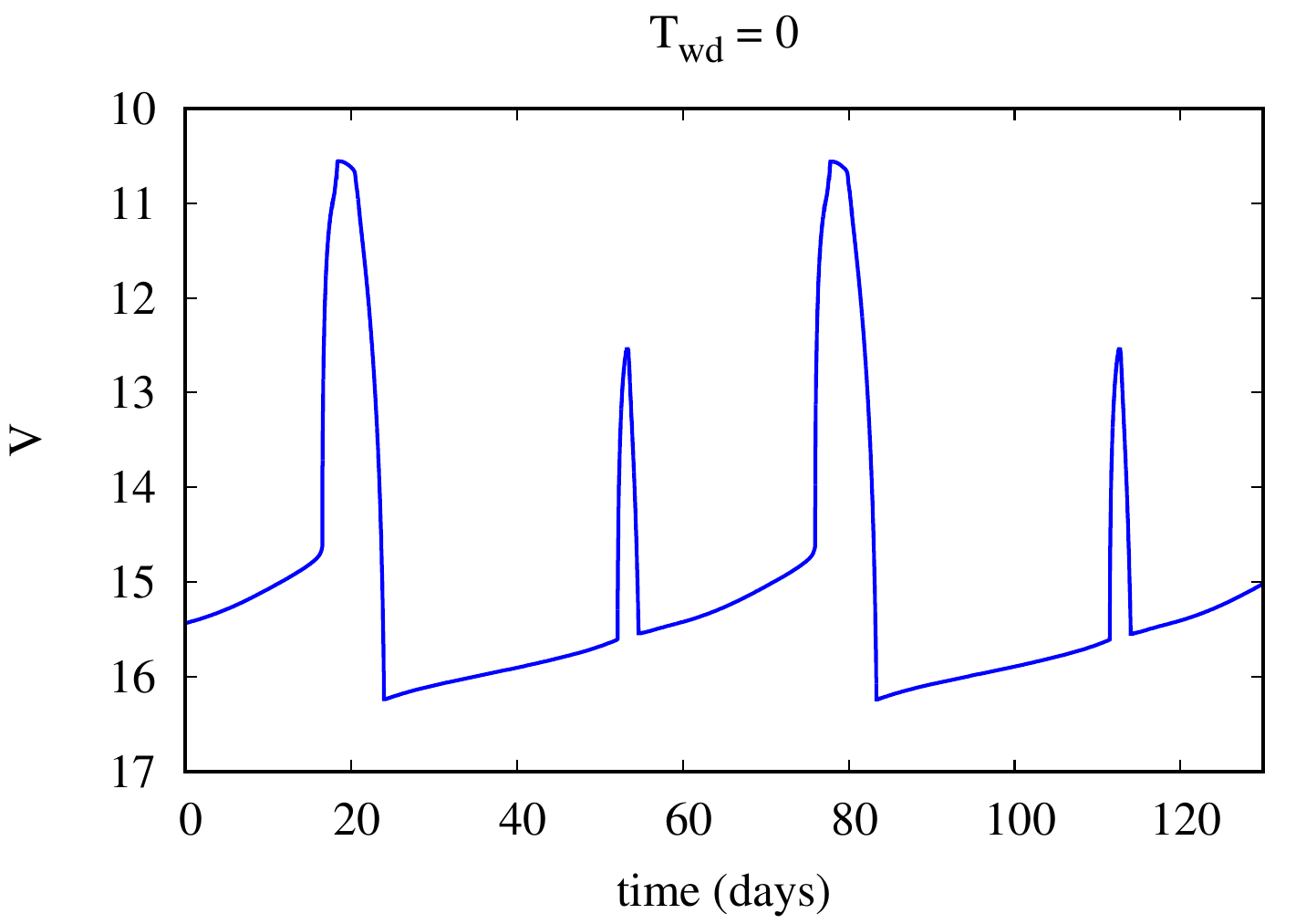}\includegraphics[width=0.33\columnwidth]{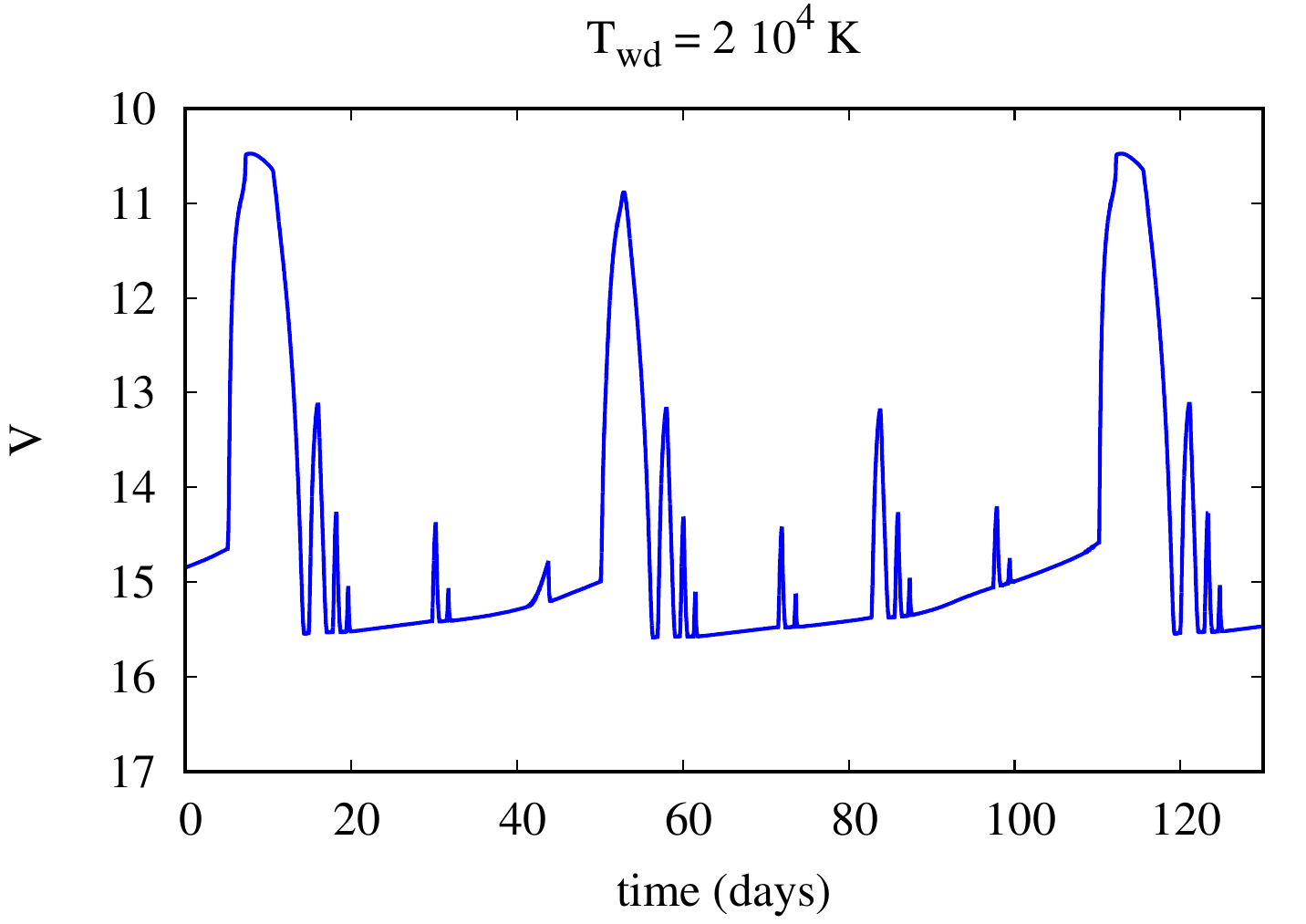}\includegraphics[width=0.33\columnwidth]{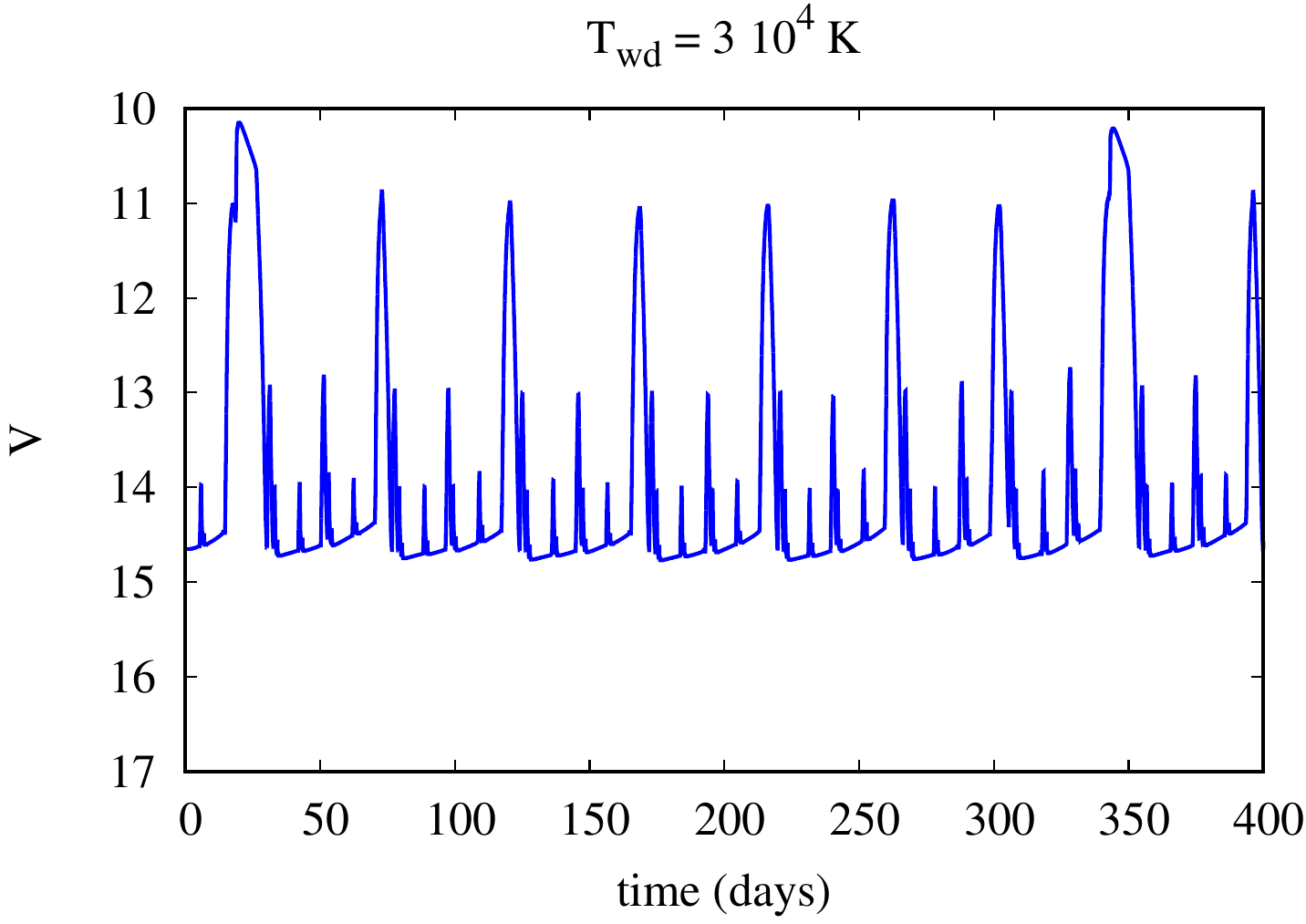}
\hspace*{0.33\columnwidth} \includegraphics[width=0.33\columnwidth]{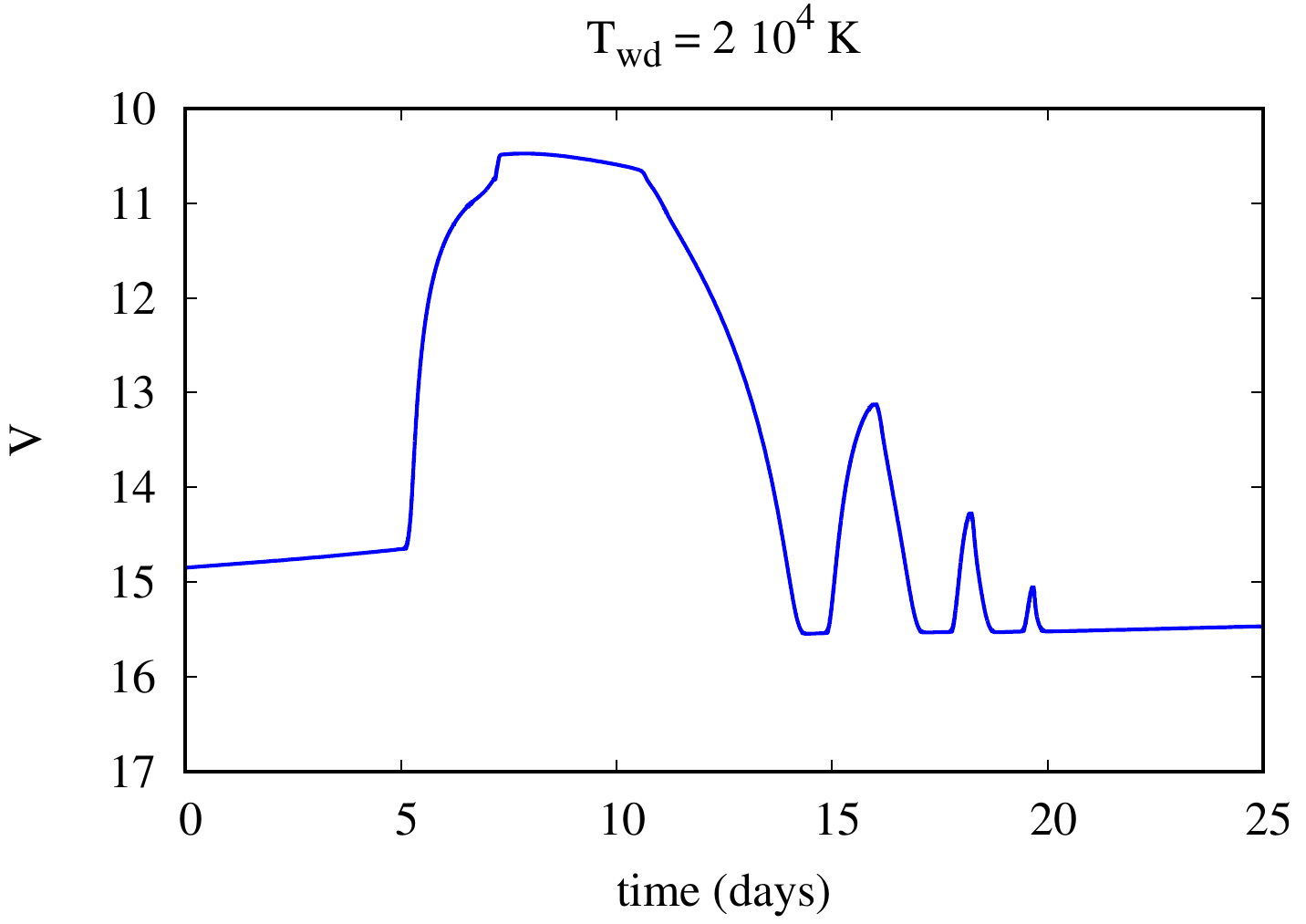}\includegraphics[width=0.33\columnwidth]{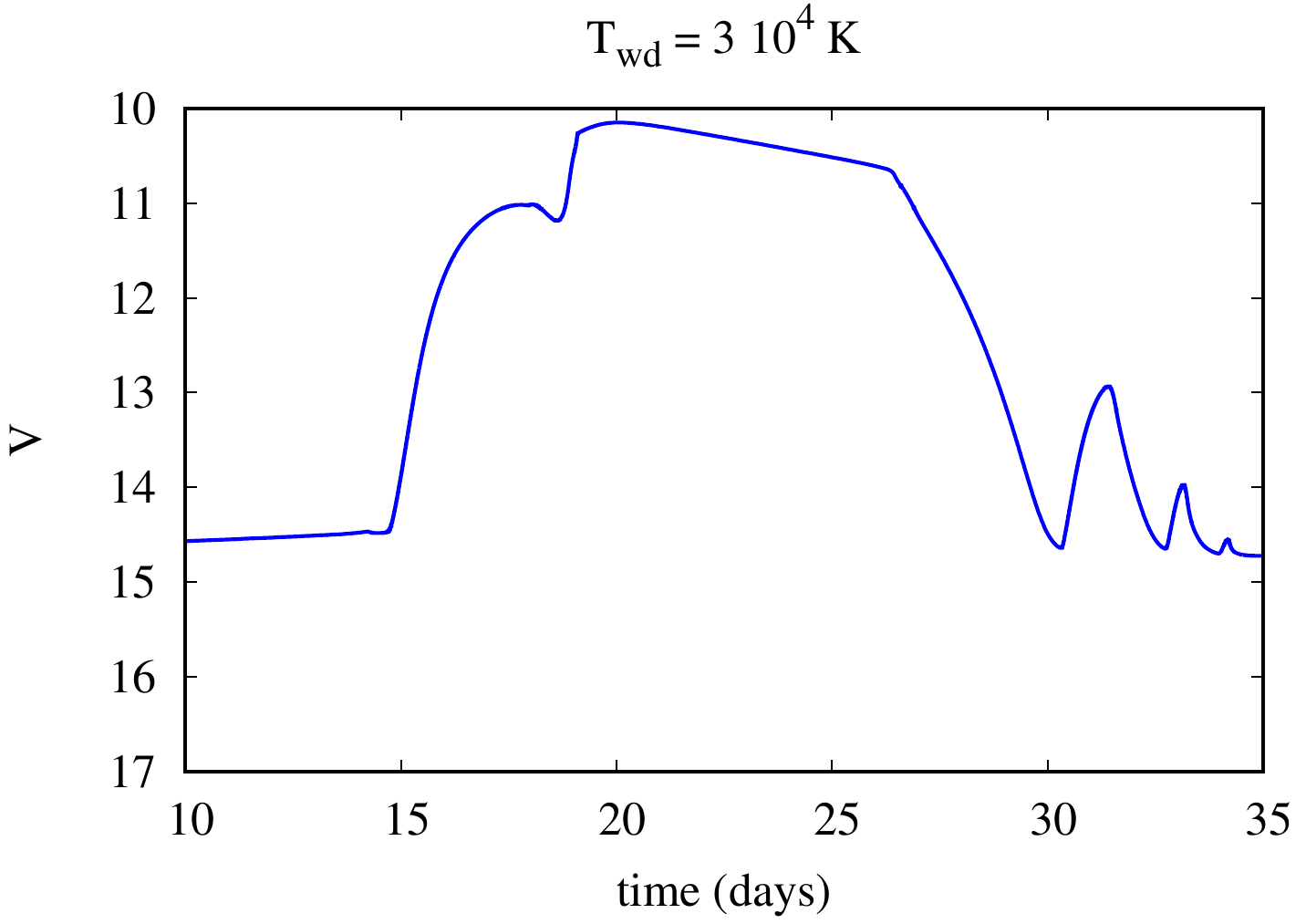}
\end{center}
\caption{Irradiation by a hot white dwarf. Here, $M_1=0.6$M$_\odot$, the orbital period is 4 hr, $\alpha_c=0.04$, and $\alpha_h=0.2$ as in Fig. \ref{fig:visc}. The disc albedo is zero. The bottom panels zoom on the major outburst, and show that the luminosity reaches its quiescent level before reflares.}
\label{fig:tstar}
\end{figure} 
 
\subsection{Irradiation by a hot white dwarf}
 
A hot white dwarf heats the disc according to \cite{f85}:
\begin{equation}
T_{\rm irr}^4 = (1-\beta) \left[ \arcsin(\rho) -\rho \sqrt{1-\rho^2} \right] \frac{T_{\rm wd}^4}{\pi}
\end{equation}
where $\rho = R_{\rm wd}/r$ is the ratio of the white dwarf radius to the radial distance, $T_{\rm wd}$ the white dwarf effective temperature, and $\beta$ the disc albedo. For $r = R_{\rm wd}$, one obtains $T_{\rm irr}^4 = (1 - \beta) T_{\rm wd}^4 / 2$, while at large distances $T_{\rm irr} \propto r^{-3}$, as for viscous heating. The ratio of the irradiating flux to the viscous flux remains thus approximately constant far from the white dwarf. Irradiation is expected to play a significant role only for very hot white dwarfs, with $T_{\rm wd} \gtrsim (1$–$1.5)\times10^4$ K, such that $T_{\rm irr} \gtrsim 10^4$ K in the inner disc.

The white dwarf continuously heats the disc with a constant flux. When the accretion rate is sufficiently high that an unirradiated disc would be marginally unstable, irradiation from a hot white dwarf can stabilize the disc on the hot branch. Otherwise, such irradiation tends to have a destabilizing effect (see Fig.~\ref{fig:tstar}). For $T_{\rm wd} = 2\times10^4$ K, the overall light curve is only moderately affected, but reflares emerge at the end of each major outburst, along with mini-outbursts during quiescence. The duration of outbursts and the recurrence time between successive major events are both increased. Notably, the disc returns to its quiescent level between reflares, an effect that, as shown in the next section, contrasts with the behaviour expected in the case of self-irradiation. For larger white dwarf temperatures, the light curve becomes increasingly complex as intermediate outbursts develop.
 
 \subsection{Self irradiation in X-ray binaries}
 
 The hot inner parts of the accretion disc can also irradiate its outer parts. The corresponding irradiating flux is given by \cite{ss73}
 \begin{equation}
 \sigma T_{\rm irr}^4 = C(r) \frac{L_{\rm X}}{4 \pi r^2}
 \end{equation}
where 
\begin{equation}
C(r)=(1-\beta) \left( \frac{H}{r} \right) \left(\frac{{\rm d} \ln H}{{\rm d} \ln r}  -1\right)
\end{equation}
and $H$ is the disc scale height. Since $H/r$ is generally larger in the inner regions than in the outer disc \citep{dlhc99}, the outermost parts of a planar disc irradiated by a central point source are predicted to remain largely unaffected, contrary to what is inferred from observations \citep{vm95}. This discrepancy implies that either the disc is not perfectly planar but warped, or that the X-ray source is spatially extended, for instance due to scattering in a corona or a wind, or both. Consequently, a constant value of $C(r)$ is often adopted as a phenomenological prescription, effectively parameterizing our ignorance of the detailed geometry. Values of $C \simeq 5\times10^{-3}$ reproduce the observed outburst properties of soft X-ray transients (SXTs) reasonably well when disc truncation is also taken into account \citep{dhl01}.

\begin{figure}
\begin{center}
\includegraphics[width=0.33\columnwidth]{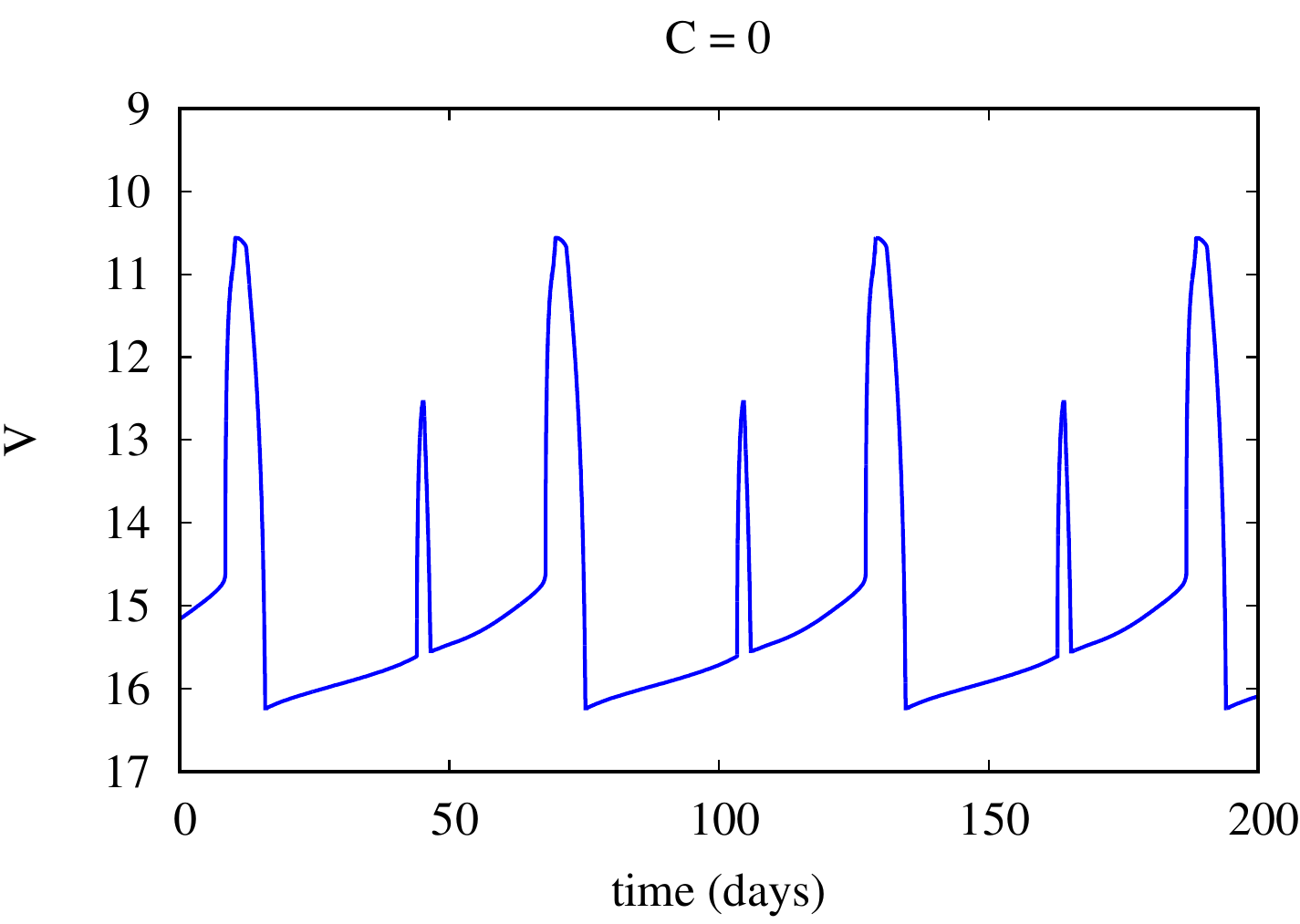}\includegraphics[width=0.33\columnwidth]{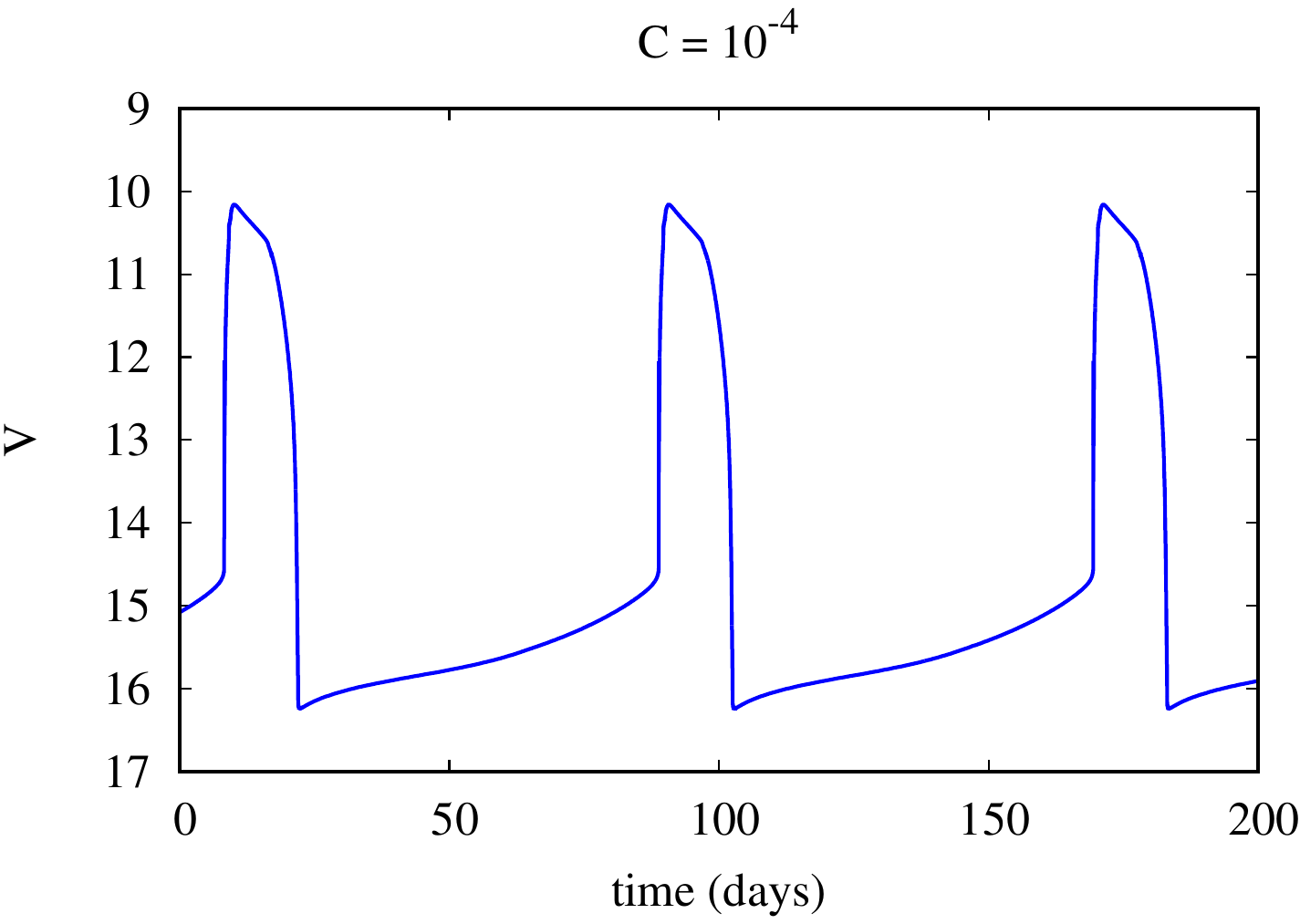}\includegraphics[width=0.33\columnwidth]{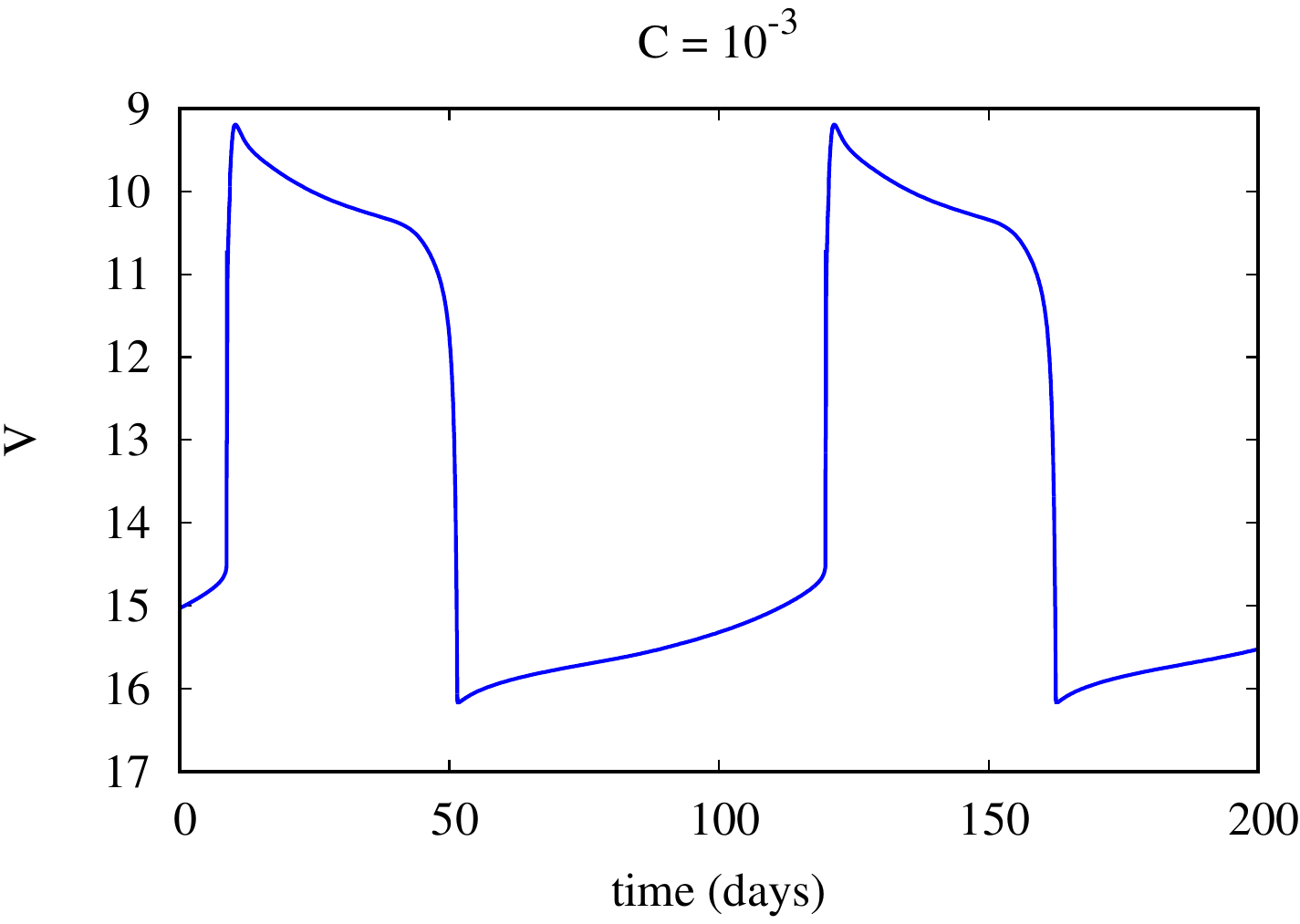}
\end{center}
\caption{Self irradiation of the disc. The binary parameters are those of Fig. \ref{fig:visc}.}
\label{fig:fill}
\end{figure}

\begin{figure}
\begin{center}
\includegraphics[width=0.40\columnwidth]{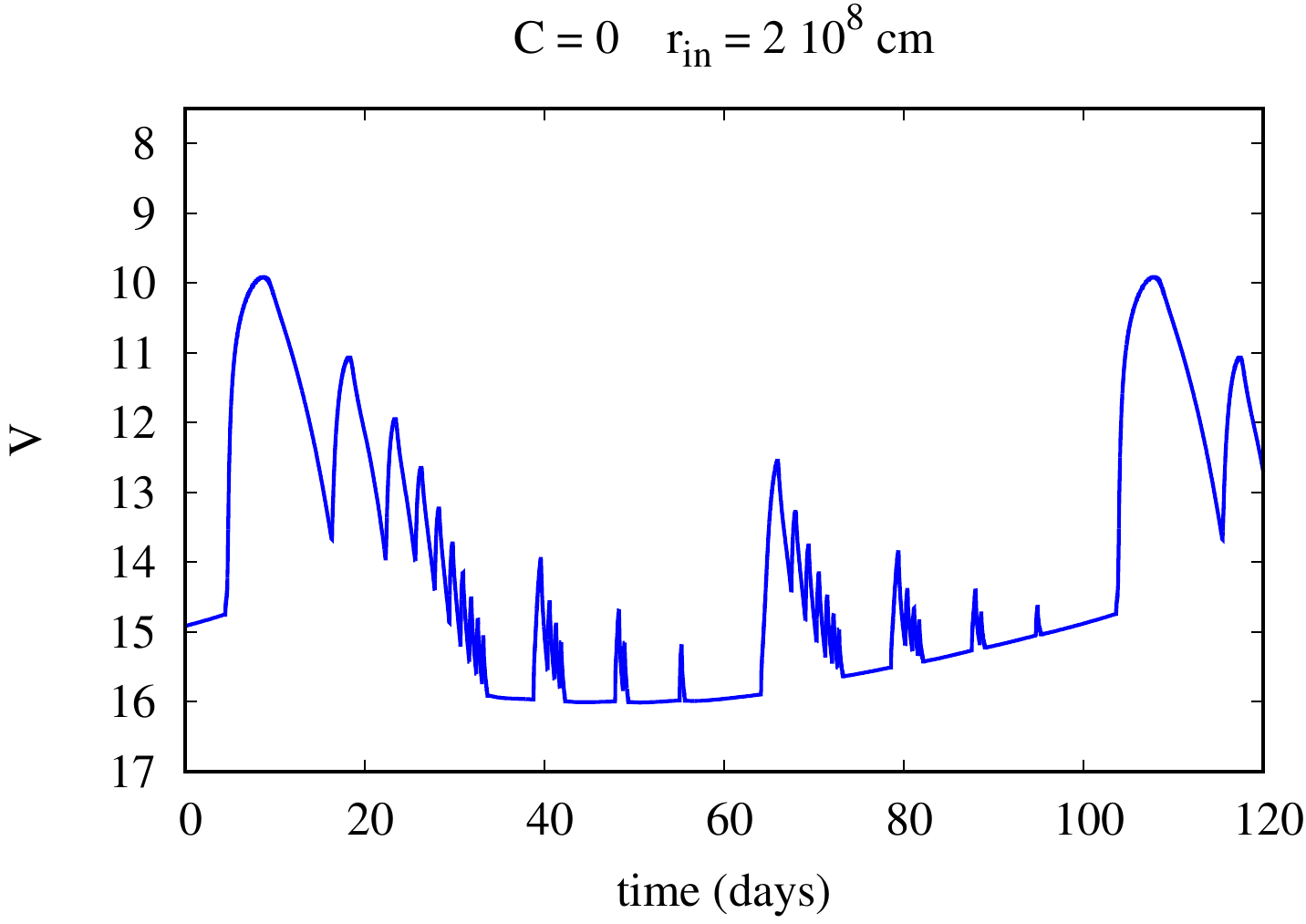}\includegraphics[width=0.40\columnwidth]{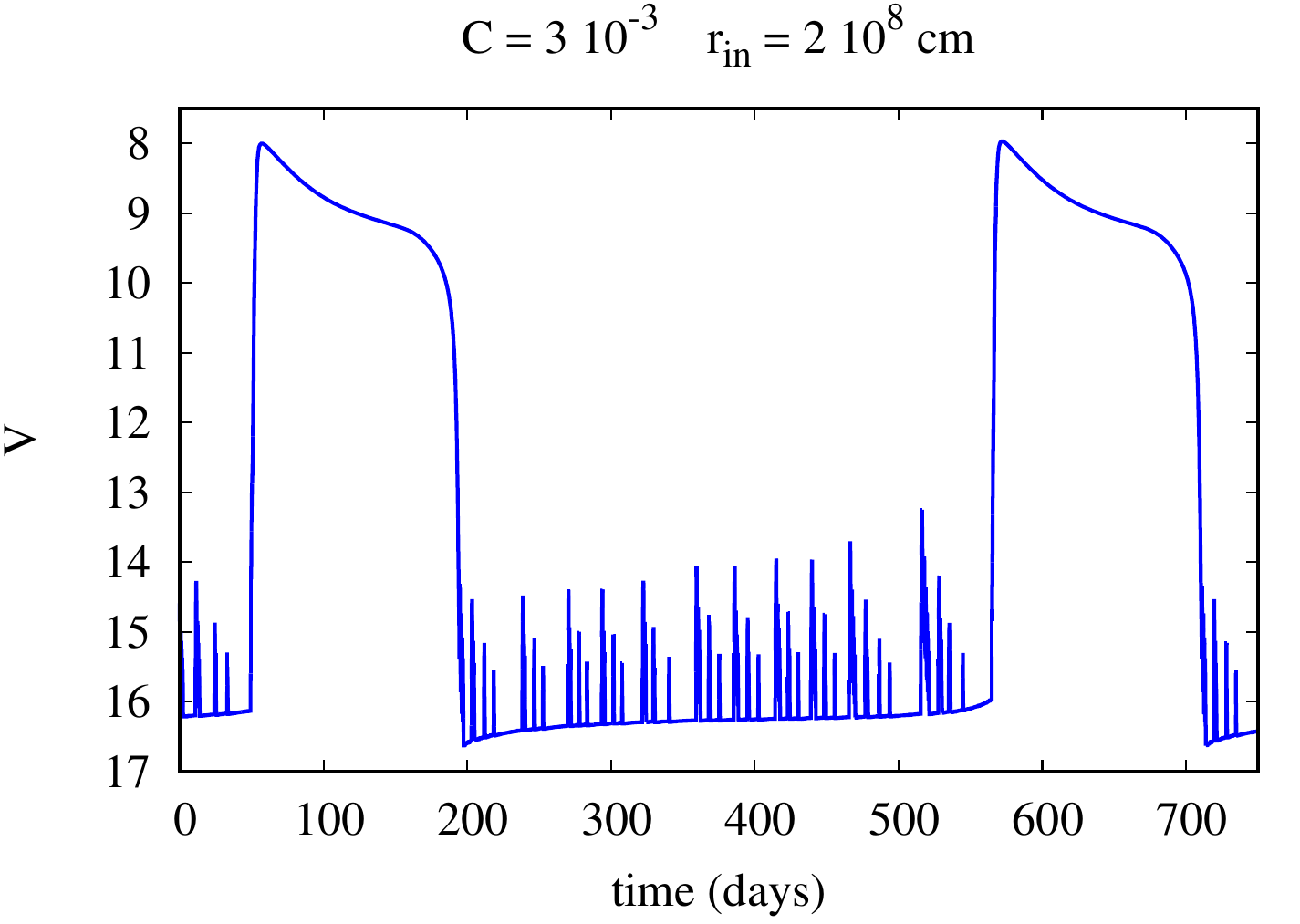}
\end{center}
\caption{Self irradiation of the disc when the inner disc radius is small. The binary parameters are those of Fig. \ref{fig:rin}.}
\label{fig:fill_rin}
\end{figure}

In contrast to irradiation by a hot white dwarf, self-irradiation of the accretion disc becomes significant only during outbursts. Figure~\ref{fig:fill} illustrates the effect of increasing values of the irradiation parameter $C$ on the model light curves. Irradiation keeps the entire disc on the hot branch for a longer duration compared to the unirradiated case, resulting in extended outbursts. As a consequence, a larger fraction of the disc mass is accreted during each event, suppressing intermediate outbursts and increasing the recurrence time.

However, this increase in recurrence time remains limited. Because the outbursts shown in Fig.~\ref{fig:fill} are of the inside-out type, the recurrence time is approximately equal to the viscous diffusion timescale. The outburst evolution can be divided into four distinct phases: an initial rise; a plateau phase during which the disc empties while remaining fully in the hot state; an irradiation-controlled decay, during which a cooling front propagates at a speed set by the irradiating flux; and finally, a thermal decay phase, when the luminosity has dropped and X-ray irradiation becomes negligible. Approximate analytical solutions reproducing these phases agree well with numerical results \citep{hl20}. It is noteworthy that the initial decay during the plateau phase, contrary to common assumptions, is not exponential but follows a $(1 + t/t_0)^{-10/3}$ dependence \citep{rk01}, where $t_0$ is a characteristic timescale.

Irradiation also tends to reduce or suppress the occurrence of reflares during the decay phase, as shown in Fig.~\ref{fig:fill_rin}. In this regime, self-irradiation produces long intervals between major outbursts, although mini-outbursts persist and recur on the diffusion timescale at the inner edge of the disc, which is relatively short. The only way to eliminate these unobserved features is to increase the viscous timescale in the innermost regions by either truncating the inner disc or by assuming a radial dependence of viscosity as in \cite{mw89}.

\section{Mass transfer variations}

\subsection{Intrinsic variations from the secondary}

Polars are cataclysmic variables in which the strong magnetic field of the white dwarf prevents the formation of an accretion disc. Their luminosity variations therefore directly trace changes in the mass-transfer rate from the secondary star. These systems exhibit pronounced brightness variations, including prolonged low states whose origin remains uncertain. It has been proposed that such low states could be caused by magnetic starspots on the secondary, located near the $L_1$ point, temporarily reducing the mass transfer rate.

Similarly, nova-like systems also show significant luminosity variations, indicating that the mass-transfer rate fluctuates on timescales of months or longer. Variations on shorter timescales are smoothed out by the accretion disc.

Mass-transfer variations are required to explain the alternation between dwarf nova and nova-like phases observed in Z~Cam systems, where the latter are referred to as standstills. This behaviour was recognized long ago, and disc instability models successfully reproduce the Z~Cam phenomenon \citep{kc98}. However, a key prediction of the models is not always satisfied: standstills should end with a decay into quiescence: IW~And-type systems exhibit standstills terminated by an outburst. It has been suggested that such events may result from a mass-transfer burst \citep{hl14} or from a tilted accretion disc allowing direct mass inflow to the inner regions \citep{kok20}, though the issue remains debated.

It has also been proposed that very short (less than one day) and weak outbursts could correspond to brief enhancements in the mass-transfer rate. Early calculations indeed showed that short outbursts are possible \citep{av89}; however, those models assumed a viscosity parameter $\alpha = 1$, which is unrealistically high. More realistic simulations \citep{hl17} indicate that, if the disc is cold, a mass-transfer burst only enhances the hot-spot luminosity, producing only a short, but very weak optical brightening. Conversely, if the disc is hot, the resulting optical outburst can reach a large amplitude, but its duration is long, of the order of the viscous timescale (i.e., several days).

\subsection{Irradiation of the secondary}

Mass transfer from the secondary can also increase in response to irradiation of its surface. For this to occur, the vicinity of the $L_1$ point must be affected by heating. Since the $L_1$ region is normally shielded from direct illumination by the accretion disc, the efficiency of this mechanism depends on whether circulation flows induced by surface temperature gradients can transport heat from irradiated to shadowed regions. Such flows, however, appear to be less efficient than initially expected \citep{vh07}. Direct illumination of the $L_1$ region may still be possible if the disc is warped \citep{c15}, or if radiation is scattered toward $L_1$ by an extended medium such as a wind or a corona.

Given these uncertainties, the effect of irradiation on mass transfer is often parametrized by assuming that the mass-transfer rate is given by
\begin{equation}
\dot{M}_{\rm tr} = \max (\dot M_0,\gamma \dot M_{\rm acc})
\end{equation}
where $\dot{M}_{\rm 0}$ is the unperturbed mass transfer rate, $\dot{M}_{\rm acc}$ the accretion rate onto the compact object and $\gamma$ a dimensionless parameter smaller than unity.

\begin{figure}
\begin{center}
\includegraphics[width=0.33\columnwidth]{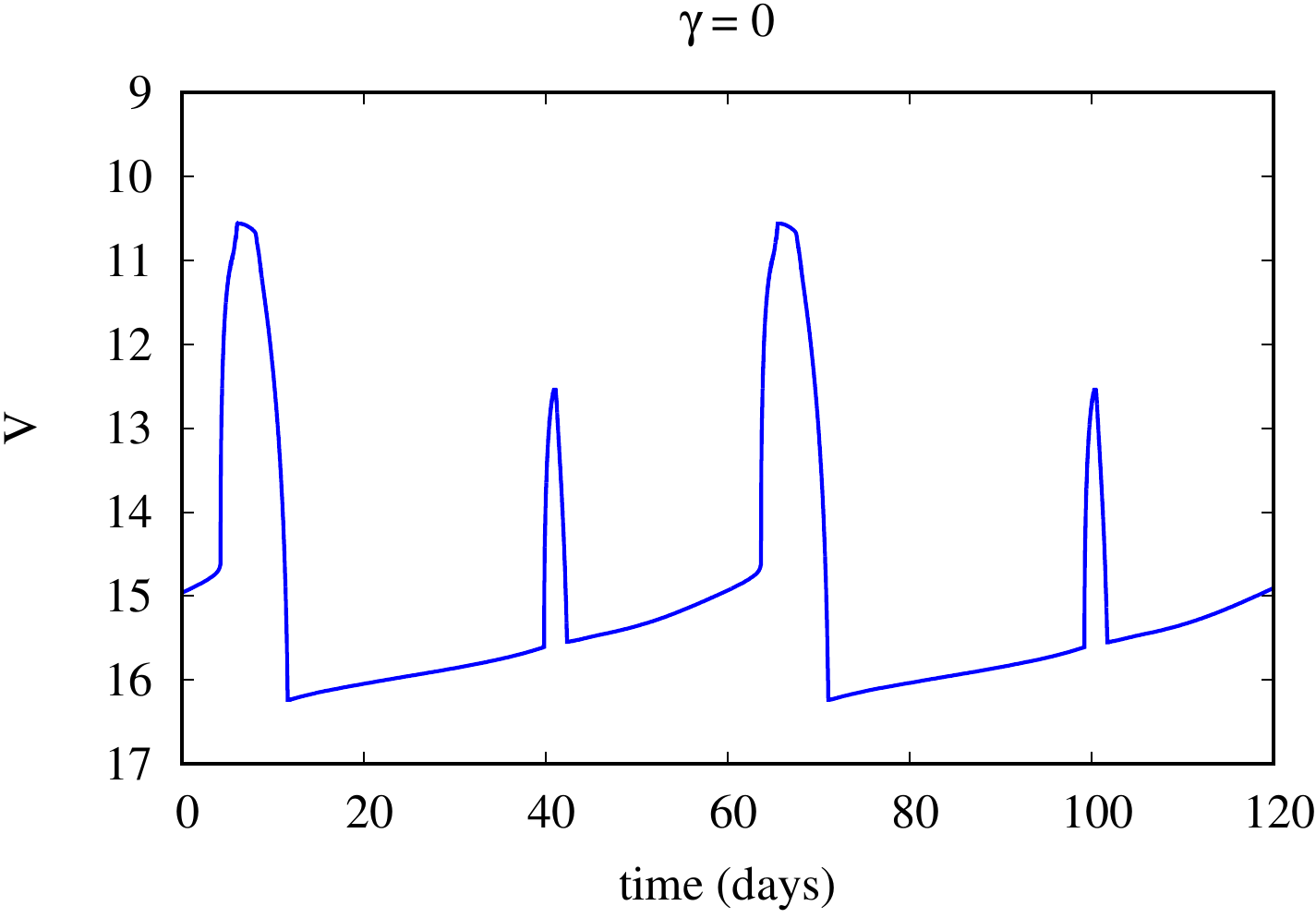}\includegraphics[width=0.33\columnwidth]{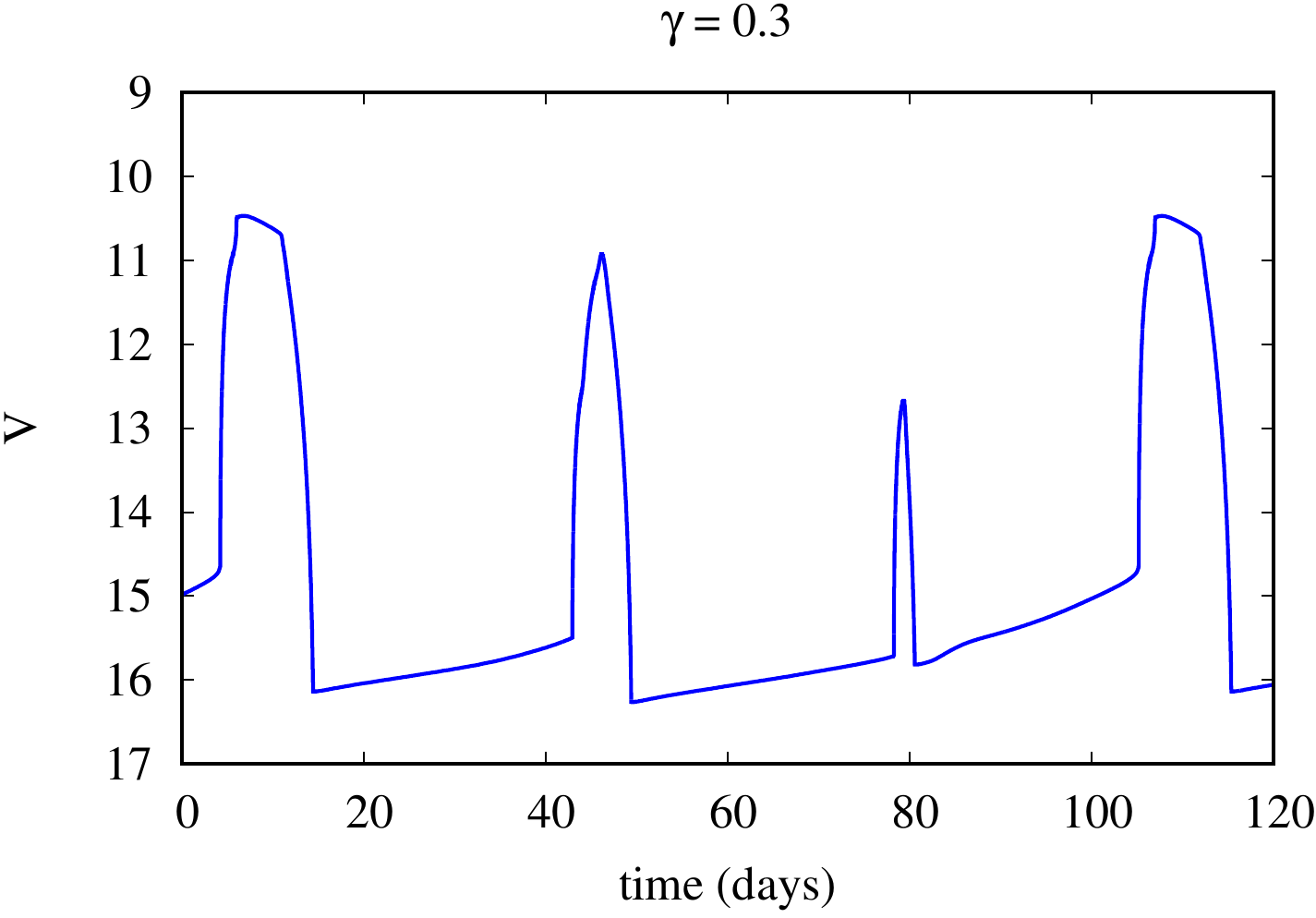}\includegraphics[width=0.33\columnwidth]{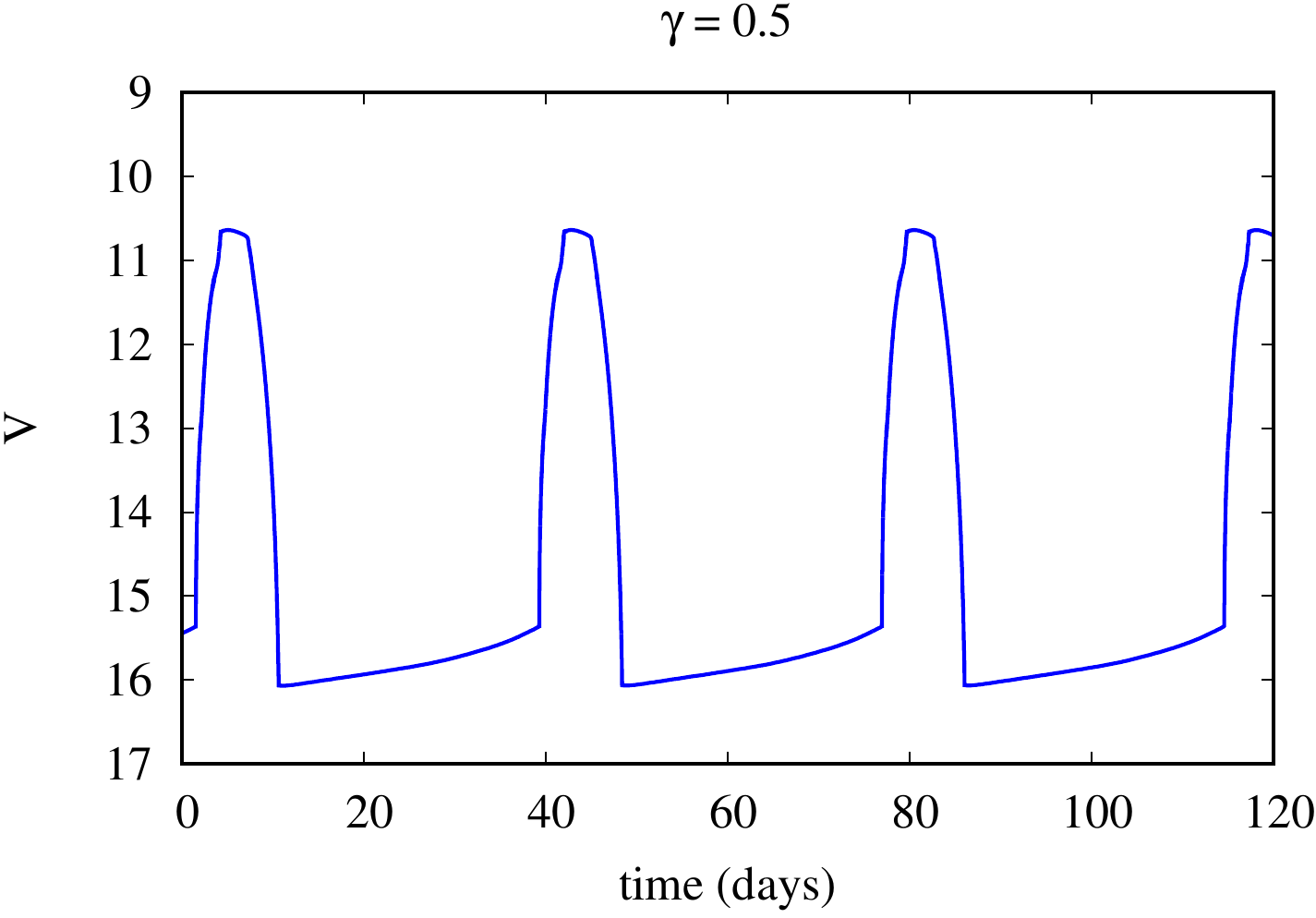}
\end{center}
\caption{Model light curves in response to the irradiation of the secondary. The binary parameters are those of Fig. \ref{fig:visc}.}
\label{fig:gamma}
\end{figure}

Figure~\ref{fig:gamma} shows light curves obtained for $\gamma = 0$, $0.3$, and $0.5$. As expected, larger values of $\gamma$ produce longer and stronger outbursts. It can also be noted that the resulting light curves may exhibit flat-topped outbursts, reminiscent of the superoutbursts observed in SU~UMa systems. It should be emphasized that the outbursts are triggered by the classical thermal-viscous instability of the accretion disc; irradiation merely modulates the mass transfer rate after the onset of an outburst, in contrast to a mass-transfer outburst where the instability originates in the secondary star.

\begin{figure}
\begin{center}
\includegraphics[width=0.4\columnwidth]{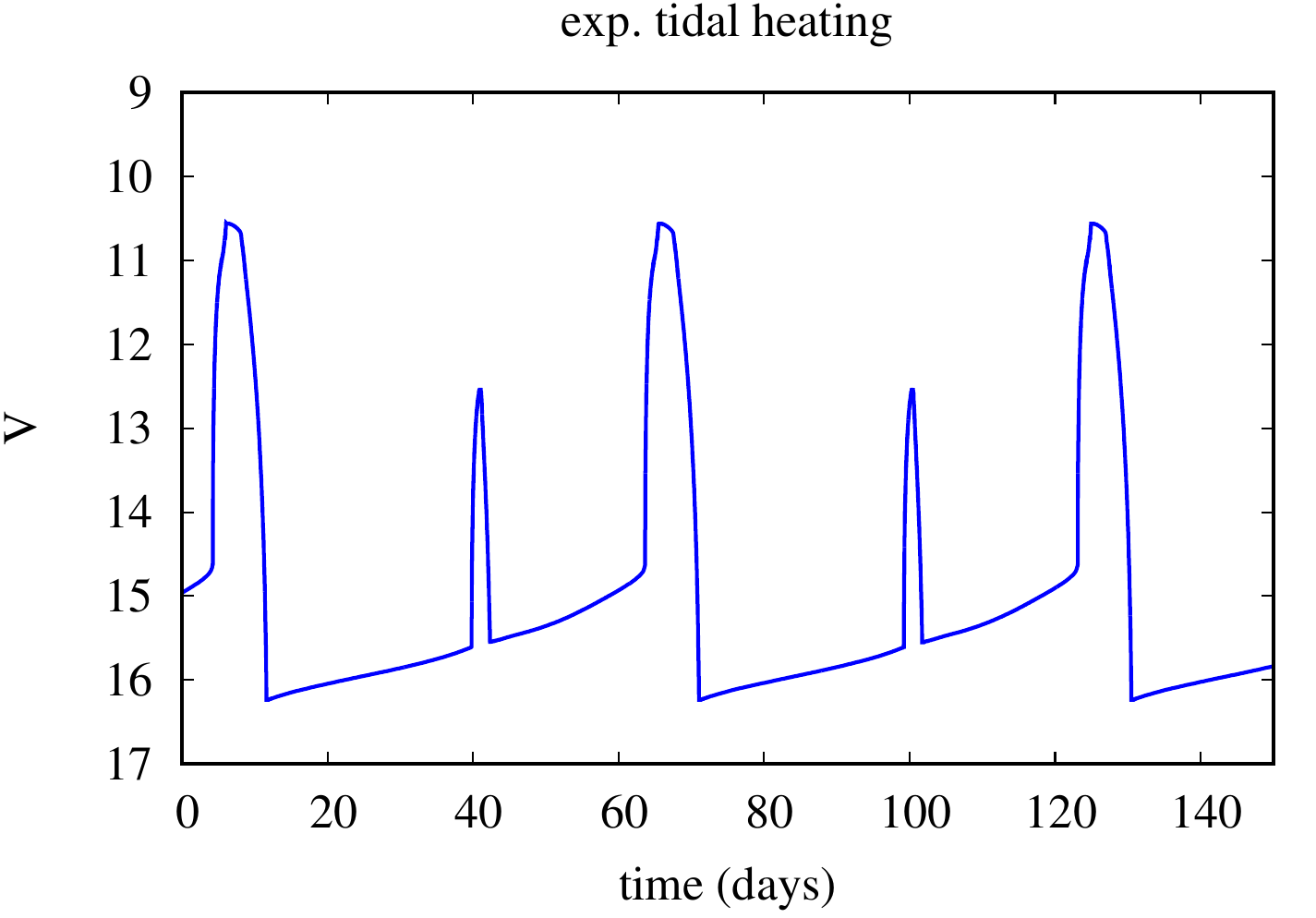}\includegraphics[width=0.4\columnwidth]{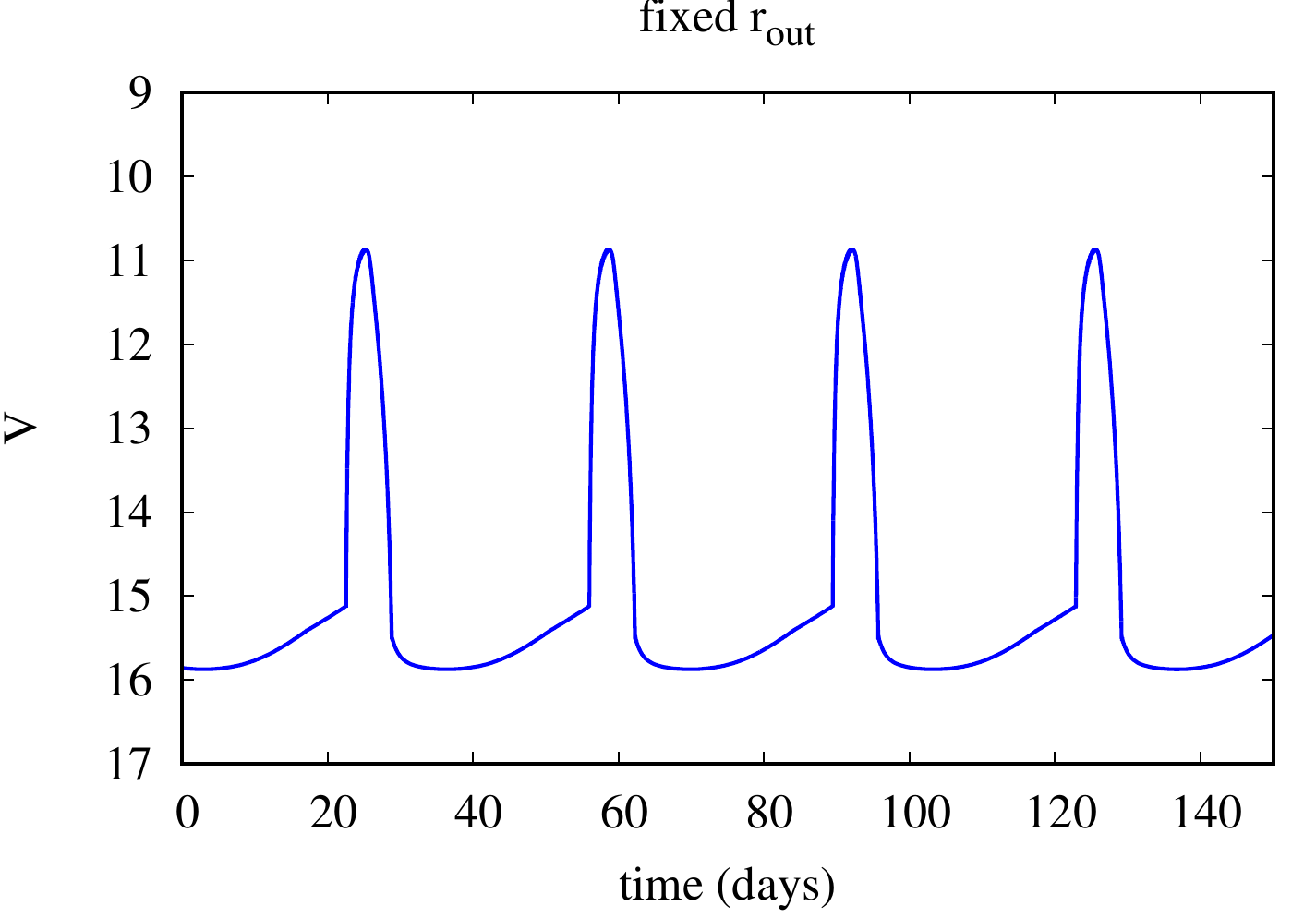}
\includegraphics[width=0.4\columnwidth]{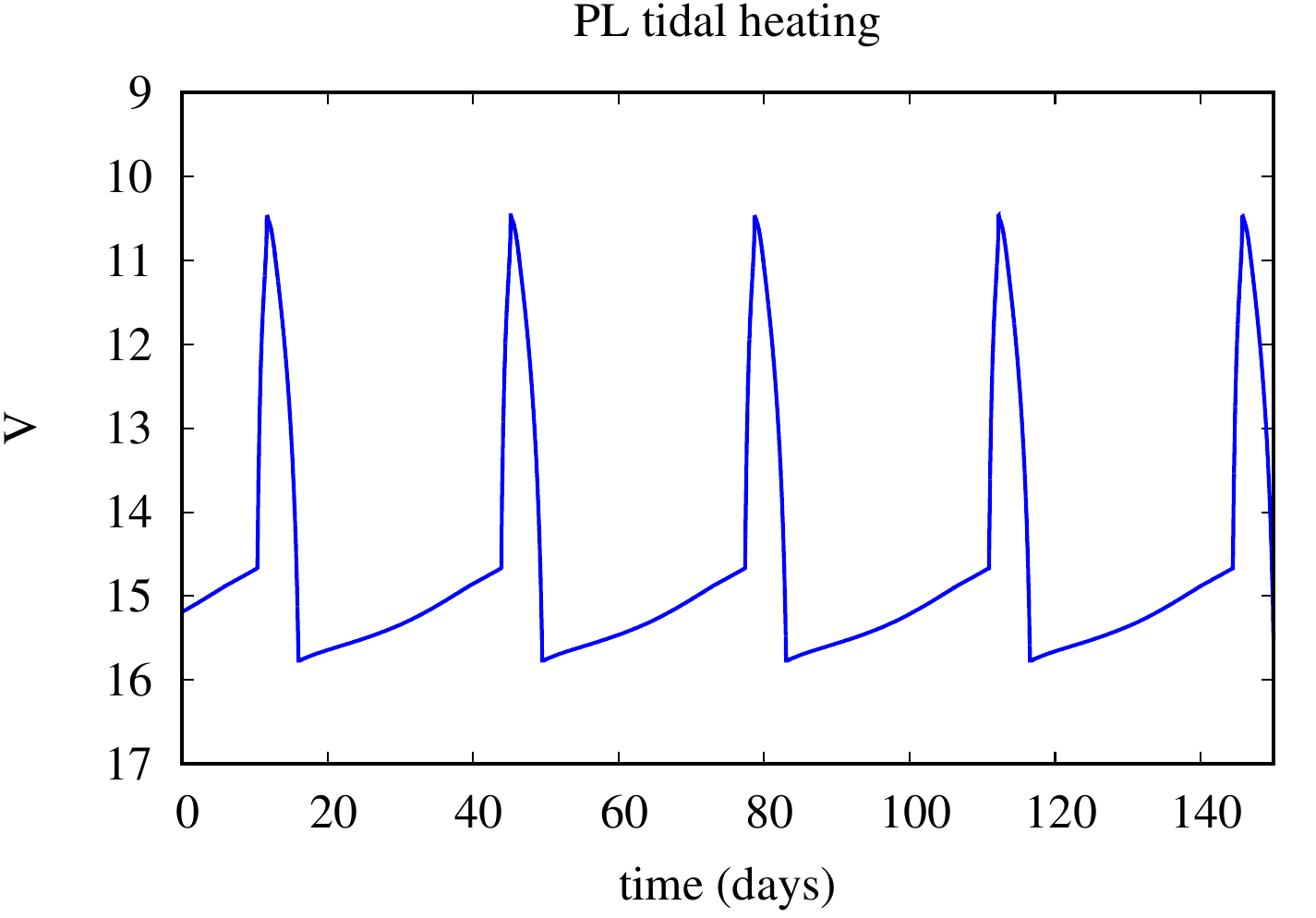}\includegraphics[width=0.4\columnwidth]{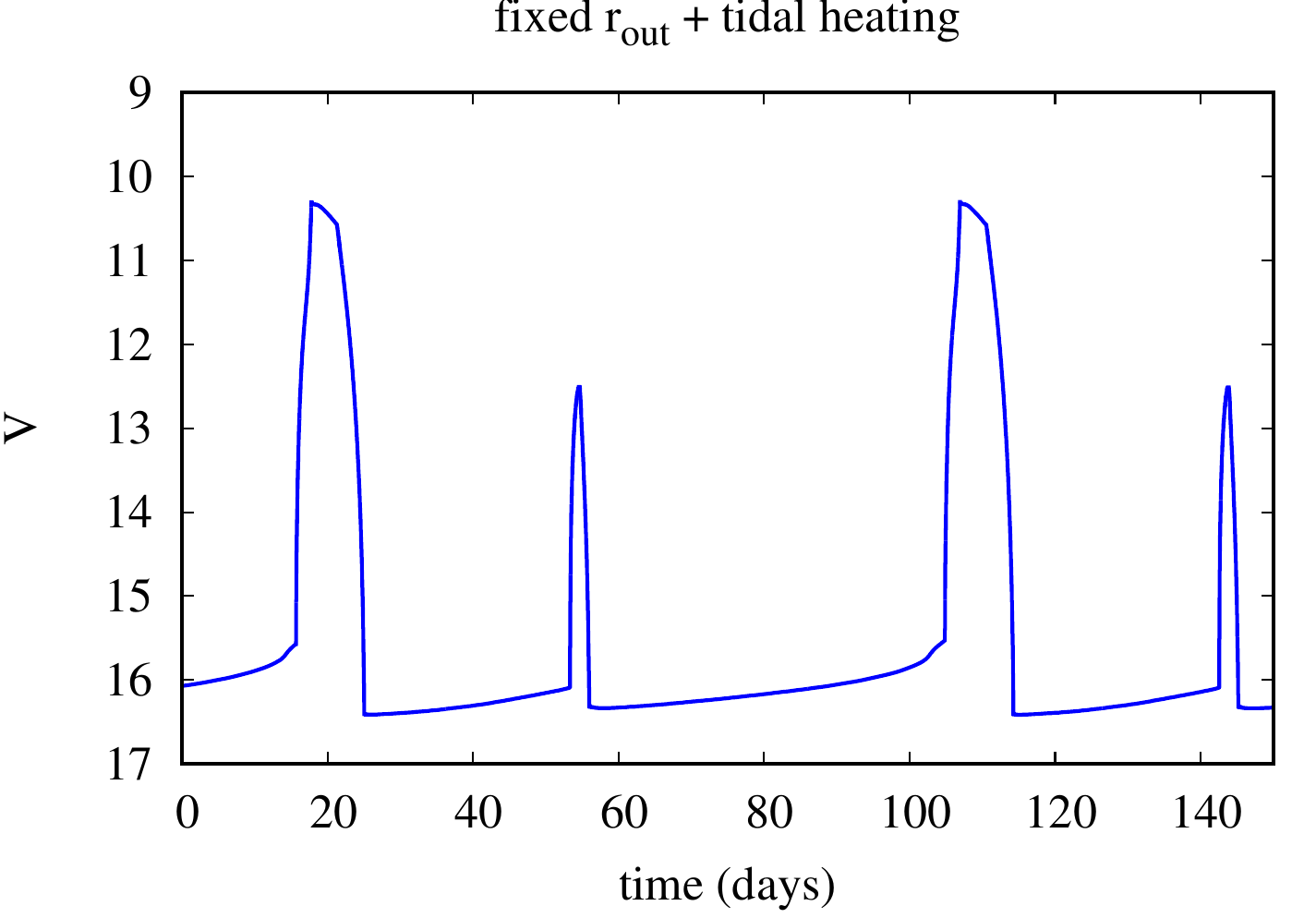}
\end{center}
\caption{Effect of various prescriptions for the tidal torque. The binary parameters are those of Fig. \ref{fig:visc}. The left panels correspond to cases where $r_{\rm out}$ is allowed to vary, whereas it is kept fixed for the right panels. Tidal heating is not included for the top right panel.}
\label{fig:out}
\end{figure}

\subsection{Tidal torques}

Significant discrepancies between model predictions have been reported depending on whether the outer disc radius, $r_{\rm out}$, is kept fixed or allowed to vary \citep{hmd98}. Moreover, the actual strength and radial dependence of the tidal torque remain uncertain. The prescription most commonly used is that of Smak \cite{s84}, based on the linear analysis of Papaloizou and Pringle \cite{pp77}, in which the tidal torque is expressed as
\begin{equation}
T_{\rm tid} = c \omega \nu \Sigma (r/a)^5
\label{eq:pl}
\end{equation}
where $c$ is a constant, $\omega$ the Keplerian angular frequency, $\nu$ the kinematic viscosity, and $\Sigma$ the surface density. The constant $c$ is usually adjusted so that, in a stationary configuration, the outer disc radius corresponds to the tidal truncation radius $r_{\rm tid}$.

However, the linear approximation breaks down as $r$ approaches $r_{\rm tid}$. Ichikawa and Osaki \cite{io94} showed that the tidal torque rises steeply near $r_{\rm tid}$, suggesting that an alternative, perhaps more realistic prescription could be written as
\begin{equation}
T_{\rm tid} = c \exp \frac{r-r_{\rm tid}}{\Delta r}
\label{eq:exp}
\end{equation}
where $\Delta r$ is an arbitrary scale parameter chosen such that $\Delta r \ll r_{\rm tid}$. 

Figure~\ref{fig:out} illustrates the impact of these different prescriptions on the model light curves. Noticeable differences are seen both in the recurrence times and in the quiescent luminosities. Although the maximum disc radius and the amplitude of its variations are similar in both torque formulations, the overall light-curve morphology differs significantly. For completeness, and because models with a fixed outer radius do not usually include tidal heating, Fig.~\ref{fig:out} presents cases where the outer radius is held constant, with tidal heating either neglected or computed using Eq.~(\ref{eq:pl}).

Tidal torques are also responsible for the disc becoming eccentric and precessing when its outer edge reaches the 3:1 resonance radius, $r_{3:1}$. This so-called tidal instability is generally invoked to explain the occurrence of superhumps and superoutbursts in SU~UMa systems, and is expected to operate in SXTs as well. A detailed discussion of this instability is clearly outside the scope of this paper. It is worth noting, however, that current models of outbursts driven by this instability \citep{o89} are typically one-dimensional: the tidal torque is arbitrarily  increased once $r_{3:1}$ is reached and maintained at this elevated level until the disc contracts well below this radius. Since the instability is intrinsically two-dimensional in nature, such simplifications raise questions regarding the reliability of these models.

\section{Stream overflow}

At high mass-transfer rates, the gas stream leaving the secondary may overflow the accretion disc \citep{ls76,h99}. It has also been proposed \citep{ko23} that such stream overflow during an anomalous event in SS~Cyg could have been responsible for the small-amplitude oscillations observed in 2021.

\begin{figure}
\begin{center}
\includegraphics[width=0.33\columnwidth]{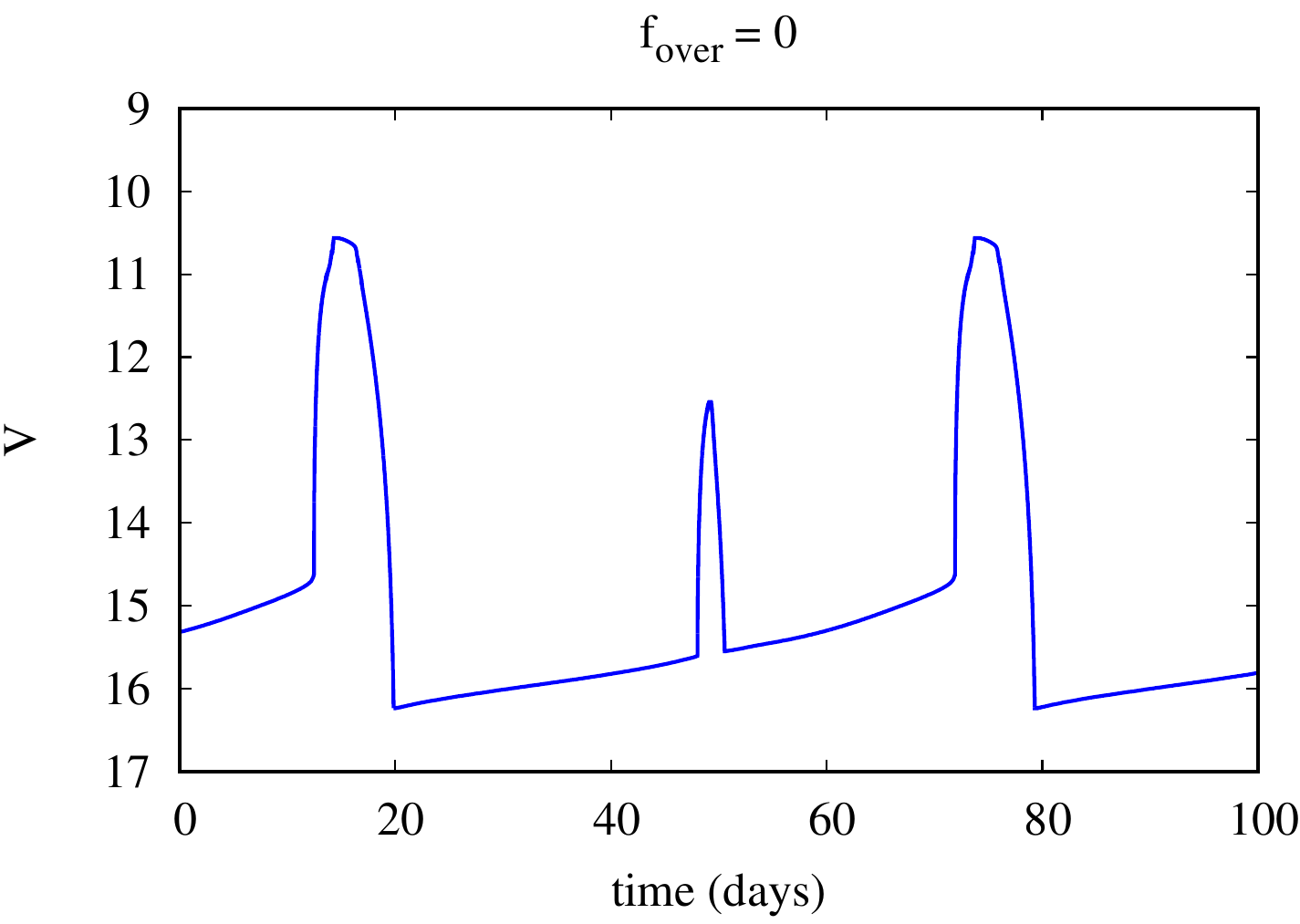}\includegraphics[width=0.33\columnwidth]{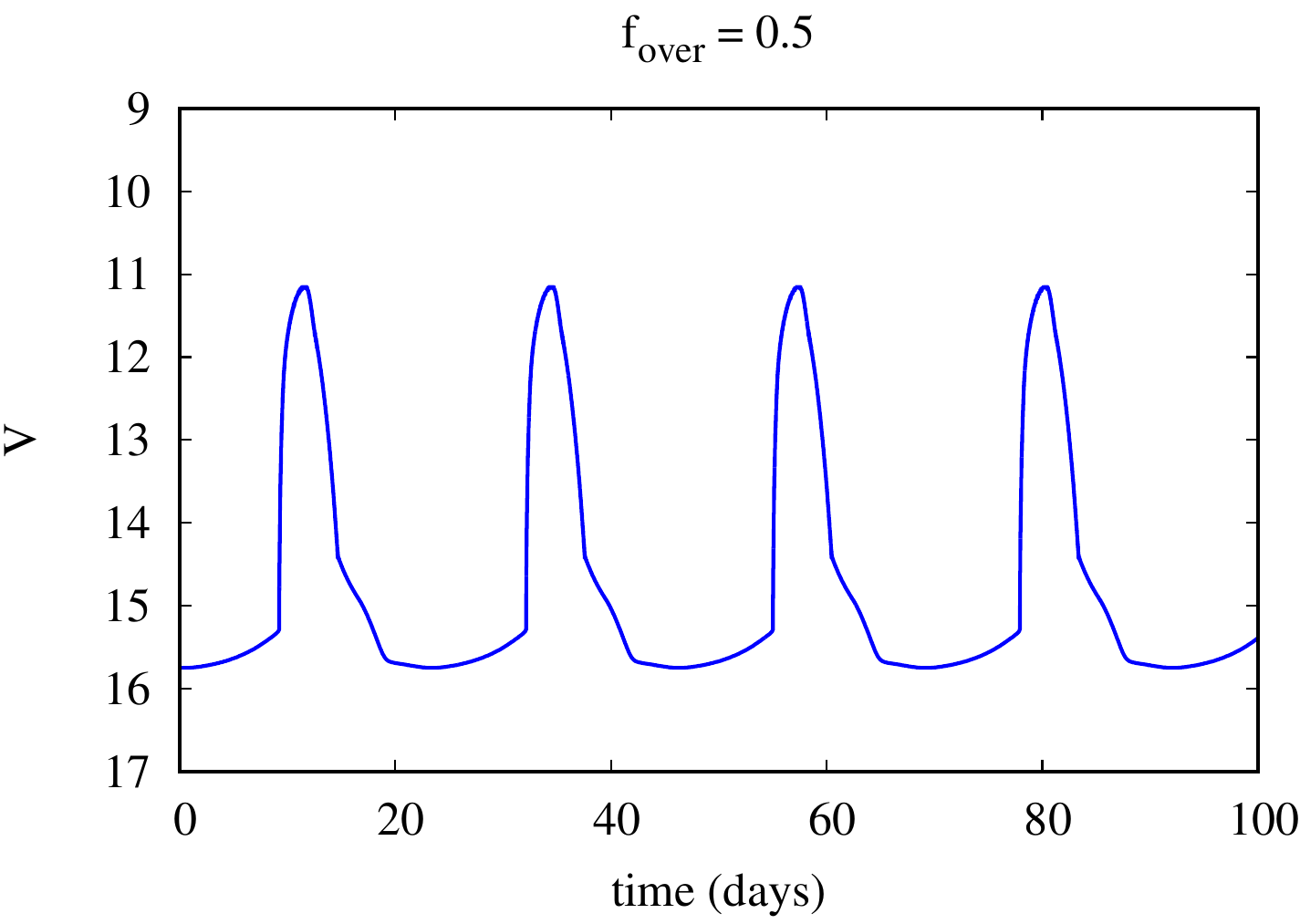}\includegraphics[width=0.33\columnwidth]{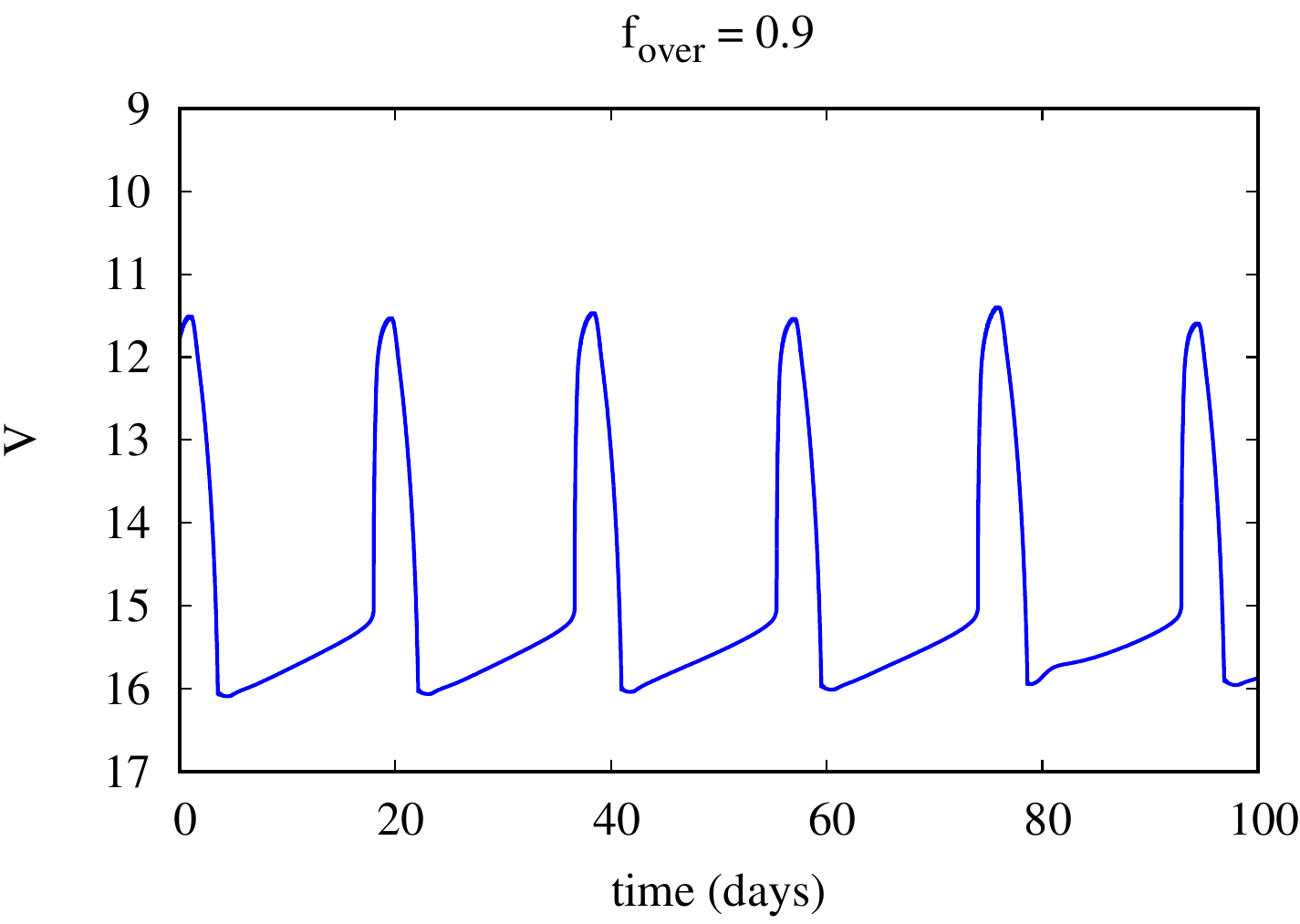}
\end{center}
\caption{Stream overflow. The binary parameters are those of Fig. \ref{fig:visc}.}
\label{fig:over}
\end{figure}

Figure~\ref{fig:over} illustrates the effect of assuming that a fraction $f_{\rm over}$ of the transferred mass is deposited at the circularisation radius, while the remaining fraction $(1 - f_{\rm over})$ is added at the outer disc edge. The energy dissipated by the stream–disc interaction has not been included, although it could be significant if $f_{\rm over}$ is large, since the detailed structure of the impact region -- and in particular the optical depth at which the kinetic energy is thermalised -- remains uncertain.

As expected, depositing mass at smaller radii shortens the recurrence time of the outbursts, since the viscous transport time toward the inner disc edge, where outbursts are triggered, is reduced.

\section{Winds}

This section is included for completeness only, as a more detailed discussion is presented in Dubus (these proceedings). Winds remove mass from the accretion disc and, if magnetized and coupled to the disc, also extract angular momentum. Their overall effect can, to some extent, be mimicked by an increase in the viscosity parameter \cite{tlh18}.

\section{Conclusions}

Current models can reproduce most of the observed behaviors, particularly when the additional ingredients discussed above are included. They can account for a wide variety of light-curve morphologies -- including the most unusual ones -- but only at the cost of introducing numerous poorly constrained parameters. In other words, the apparent success of the DIM in matching observations should not be taken as proof of the physical validity of all the ingredients that have been incorporated into the model. It is worth recalling, for instance, that long outbursts can be explained either by invoking tidal instabilities or by an increase in the mass-transfer rate during outbursts caused by irradiation of the secondary. What is now required is a better inclusion of realistic physics in the models.

Two decades ago, there was hope that full 3D MHD simulations would resolve these issues and render 1D codes obsolete, but this has clearly not yet been achieved. In the shorter term, progress will likely come from 2D simulations that relax the assumption of axisymmetry while retaining accurate vertical structures, as in \cite{jwk24}. This does not mean that 1D models will become irrelevant, given their computational efficiency, but they can be benchmarked against 2D results to assess the parametrizations they rely on and to constrain some of the presently free parameters.

Several outstanding problems remain, which may require substantial modifications to the DIM. The reader is referred to \cite{l23} for a discussion of some of these issues. Among the most pressing is the nature of the low states. The MRI becomes inefficient at low temperatures, especially in systems with long recurrence times such as SXTs or WZ~Sge stars. Yet accretion persists during quiescence, indicating that angular momentum transport remains effective. Another key question concerns the physical mechanisms responsible for disc truncation. Finally, the role of magnetic fields -- not only on the small scales relevant to MRI, but also on larger scales that may affect angular momentum transport and have direct observational signatures in the spectra -- must be better understood \cite{np19,znn24}.

\acknowledgments
I thank the Fujihara foundation for financial support which enabled me to attend the workshop.

\bibliographystyle{JHEP}
\bibliography{biblio}

\end{document}